\documentstyle[11pt,aasms4]{article}
\def\etal{{\it et al.\ }}
\def\eg{{\it e.g.,}}
\def\ie{{\it i.e.,}}

\def\Lya{Ly$\alpha$}

\begin{document}

\title{FAUST observations of UV sources toward the Virgo cluster}
\author{Noah Brosch
\thanks{E-mail: noah@wise.tau.ac.il}, Liliana Formiggini\thanks{also at the Istituto di
Radioastronomia del C.N.R., Bologna, Italy}
\thanks{E-mail: lili@wise.tau.ac.il} \, and Elchanan Almoznino\thanks{E-mail:
nan@wise.tau.ac.il}  \\ The Wise Observatory and 
the School of Physics and Astronomy \\ Raymond and Beverly Sackler Faculty of Exact Sciences \\
Tel Aviv University, Tel Aviv 69978, Israel 
\\ \and Timothy Sasseen \thanks{tims@ssl.berkeley.edu}, 
Michael Lampton \thanks{mlampton@cea.Berkeley.EDU} and Stuart Bowyer
\thanks{bowyer@ssl.berkeley.edu} \\
Space Sciences Laboratory and Center for EUV Astrophysics \\ University of California 
\\ Berkeley CA 94720, U.S.A.}

\begin{abstract}
We analyze three UV images covering a $\sim100$ square degree field toward the Virgo cluster,
obtained by the FAUST space experiment. 
We detect 191 sources to a signal-to-noise ratio of 4.4 and identify 94\% of
them. Most sources have optical 
counterparts in existing catalogs and about half are 
identified as galaxies. Some sources with
no listed counterpart were observed at the Wise Observatory. We 
present the results of low resolution visible spectrophotometry and discuss
the foreground 101 stellar sources and the 76 detected galaxies, both in the
cluster and in the fore- or background. We derive conclusions on star-formation
properties of galaxies and  on the total UV flux from discrete and diffuse sources in the
cluster. We test for the presence of intra-cluster dust, determine the clustering properties of 
UV emitting galaxies, and derive the UV luminosity function of Virgo galaxies.
\end{abstract}


\section{Background}

The ultraviolet (UV) segment of the spectrum is an excellent tool to identify 
and quantize active star
formation (SF) regions (Donas \etal 1987). It is also a sensitive probe of 
evolved hot stellar populations (\eg \, Burstein \etal 1988). Therefore, UV 
observations may serve 
to understand processes of SF and stellar evolution. These processes are amplified and 
enhanced in a cluster environment, due to interactions among galaxies, and between galaxies and the
intracluster medium. In the case of the Virgo cluster (VC) the observations are optimized by the
proximity of the region, despite the VC not being a fully relaxed cluster. Throughout this work we 
{\bf assume} a uniform distance of 18 Mpc to all the galaxies in the VC. A smaller cluster distance, 
$\sim14$ Mpc, is suggested by the [OIII] distances to planetary 
nebulae (Jacoby \etal 1990), but Visvanathan and Griersmith (1979) 
give 17.9 Mpc, in agreement with Sandage and Tammann (1974). We note that
the depth of the cluster ($\sim$4 Mpc) is significant in comparison to the distance to the cluster.

The VC was studied intensively by many groups; we shall not give here a
thorough review of all publications but mention in particular the extensive compilation 
of Binggeli, Sandage and Tammann (1985, BST). Their VC catalog relied mainly on the morphological
appearance of a galaxy on large plate-scale photographs, to determine whether an object is member 
of the VC or lies in its fore- or background. In addition, the many papers by Hoffman and collaborators
(Hoffman \etal 1989a, 1989b; Hoffman 1989; Hoffman \etal 1995) used HI observations from the
Arecibo Observatory to determine the large-scale-structure in the direction of the VC. The combination
of HI and optical redshifts for many of the galaxies indicates the existence of a void in
the distribution of galaxies located behind the VC and in front of the Great Wall at 
5500$\leq$cz$\leq$13500 km s$^{-1}$.

The complex structure of the VC region was analyzed by Hoffman \etal (1989a) in the context
of the blue compact dwarf galaxies (BCDs) in the cluster. Their Fig. 4 emphasizes the positions
of the various groups and clouds of galaxies in this region.
They also established upper limits to the number of HI clouds devoid of optical counterparts
in the cluster, and considered other observational properties of these galaxies,
such as their FIR emission from IRAS and optical properties. Hoffman \etal (1995)
studied the background galaxies beyond the VC, in the context of their clustering properties.

A reanalysis of the velocity distribution of VC galaxies (Binggeli, Popescu and Tammann 1993)
showed a slight difference in the cluster redshift (1050$\pm$35 km s$^{-1}$) from the
previously accepted value (1094 km s$^{-1}$; Binggeli, Tammann and Sandage 1987). In addition to the
void behind Virgo mentioned by Hoffman \etal (1995), they found a significant
lack of galaxies behind Virgo, from 3000 to 3500 km s$^{-1}$. 
We consider here all galaxies in the direction of the VC and with v$_{\odot}<3000$ km s$^{-1}$
as cluster members, disregarding their membership in the Virgo core or the various sub-clusters.

X--ray observations with the ROSAT PSPC (Trinchieri \etal 1994) showed that at least 
one Virgo cluster galaxy
(NGC 4636) may be interacting with VC gas to produce significant X-ray flux at $\sim$3 Mpc from M87.
Other such galaxies may be M86 (Forman \etal 1979) and possibly NGC 4472 (Trinchieri \etal 1986).
In addition, we mention the discovery of very soft X-ray emission from M87 (Lieu \etal 1996)
by the EUVE satellite,
with 31 counts per 1000 sec in the 100\AA\, band. The X-ray and far-infrared emission from
M87, together with subtle optical asymmetries seen mainly in the blue, induced White \etal
(1991) to propose that this galaxy is stripped of its interstellar material by some hot gas, 
present at the core of the VC.

High resolution HST observations of \Lya \, absorptions in the UV spectrum of 3C273 
at the redshift of the VC
(Weymann \etal 1995 and references therein) indicate that the absorptions are probably produced by 
low-column-density clouds with low metallicity. One need they clearly identify is to evaluate
the metagalactic background ionizing radiation field, away from the vicinity of normal galaxies; we 
attempt this by using new UV observations from approximately the entire Virgo cluster, obtained
by the FAUST payload.

\subsection{Previous UV observations}

A few galaxies in the Virgo cluster were observed in the vacuum UV by the ANS satellite
(Wesselius \etal 1982). More were 
observed by the rocket payloads of Smith and Cornett (1982) and Kodaira \etal (1990), and recently 
by the FAUST imager 
(see section 2, below) 
on board the ATLAS--1 Shuttle experiment (Bowyer \etal 1993). Additional 
measurements from a balloon--mounted telescope were reported by Donas \etal (1987).

Smith and Cornett (1982) observed the cluster with
a wide field camera (11$^{\circ}.4$ FOV) with a CsTe intensifier 
coupled with photographic film behind a quartz window, which define a passband with 
FWHM$\simeq$1150\AA\, and constant--energy mean wavelength of 2421\AA\,. 
Their absolute calibration relies of $\rho$ Vir, calibrated during the 
TD--1 mission against Vega, and UV magnitudes or upper limits are given for 201 galaxies.

Donas \etal (1987) used the SCAP 2000 balloon--borne telescope to image various regions of 
the sky in a band $\sim$125\AA\, wide centered near 2000\AA\,. The SCAP telescope was a 17 cm
diameter reflector with a 6$^{\circ}$ field-of-view, imaging on film behind a UV-to-visible
image converter/amplifier. Among the galaxies listed by Donas \etal 
12 are members of the Virgo cluster.

The GUV experiment (Onaka \etal 1989) consisted of two 17 cm diameter imagers with a 
4$^{\circ}$ field of view and was launched in 1987 on board a 
sounding rocket. The GUV sensitivity response had a FWHM$\simeq$200\AA\, and peaked near 1450\AA\,. 
The results relevant to Virgo cluster galaxies were presented by Kodaira \etal (1990) and 
a discussion of the diffuse UV light (mainly galactic) in the direction of the Virgo cluster 
was given by Onaka and Kodaira (1991). The galaxies' paper gives UV fluxes for 42 galaxies 
observed by GUV and upper limits for another 47 galaxies. Note that the reported final
angular resolution of the GUV experiment was 16'.0$\times$8'.2 (Kodaira \etal 1990) and the
equivalent exposure time on the Virgo cluster was 176 sec.

\section{FAUST observations of the Virgo cluster}

FAUST
 (FAr Ultraviolet Space Telescope)
is a wide-field ($\sim8^{\circ}$ field of view) telescope with a bandpass between 1400 
and 1800\AA\ and an angular resolution of 3.5 arcminutes.  It utilizes a microchannel plate 
detector with wedge and strip anode, which records the position of each detected photon. 
Descriptions of FAUST and its operation aboard ATLAS--1 were given by Lampton \etal (1993) 
and Chakrabarti \etal (1993). Details of the image construction and subsequent reductions 
are given in Bowyer \etal (1993). A catalog of 4660 UV sources detected by FAUST was produced
by Bowyer \etal (1995).

The Virgo cluster region was imaged on three different FAUST frames, with typical exposures 
ranging from 630 sec (Northernmost frame) to 1250 sec (Southernmost frame). The images are 
separated by $\sim4^{\circ}$ in declination, with the middle image centered on M87 and each pixel 
is $\sim$1.11 arcmin. The  point source fluxes 
can be converted to physical units with a calibration derived in flight from observations of 
stars that were observed by IUE (Bowyer \etal 1993). The total sky coverage of the three Virgo 
cluster frames, defined as the region with non--zero exposure and allowing for image overlap, 
is $\sim100$ square degrees. The entire region is at high galactic latitude 
(b$\geq$65$^{\circ}$) and is shown in Figure 1.
A general discussion of UV properties of some galaxies observed 
by FAUST, in which 29 Virgo cluster objects are included, was given by Deharveng \etal (1994).

An impartial source detector algorithm, described in Brosch \etal (1995), operated on the 
three FAUST images of the Virgo cluster and produced a list of 191 UV sources, some appearing 
on more than one image, with their FAUST UV flux and equatorial coordinates (Table 1). 
Objects for which the coordinates were calculated from an astrometric solution based on
bright, well-identified stellar images, have their coordinates in Table 1 marked with an asterisk.
Note that, because of the character of the FAUST images, different regions of the images are
exposed for different times. Therefore, the detection of sources here is not
flux-limited, but rather signal-to-noise limited. The flux was measured with a simulated round
aperture of diameter 13.2 arcmin. This is large enough to include all the flux from any
galaxy in the VC, thus we did not distinguish between point and extended sources. A few very
bright UV sources (stars brighter than $\sim$6.5) were measured with a larger simulated
aperture. The sky contribution was derived from a square box with 44 arcmin sides, except for the
few bright sources, where the sides were 77 arcmin.

Table 1 lists also the  
error in the flux; this was calculated by combining in
quadrature the instrumental and systematic errors. The first
was obtained from the ``photon statistics'' distribution
of the events from which the sky brightness was calculated. 
The
systematic errors, originating from the laboratory calibration and the
intrinsic errors in the IUE scale, are estimated 
by  Bowyer \etal (1993) to be 15.8\% of the measured flux, {\it i.e.,} an additional error of 0.16 mag to
the monochromatic magnitude used here. 
The  full analysis of the UV source list, the identification of UV sources with catalogued 
objects, and follow-up optical observations of non-catalogued sources, are  
the subject of this paper. 

\subsection{Internal consistency of the FAUST measurements and source identification}

The three FAUST images have considerable overlap. This allows the comparison of 
magnitudes for a significant number of sources, which appear on two or three images.
The comparison of different 
 magnitudes
for the same objects is shown in Figure 2.
Objects measured three times, which appear in the region of overlap of all images,
are represented by three points
connected with straight lines,
 each corresponding to one pair of measurements.
Fig. 2 shows 133 pairs of values, after deleting the six brightest and
the three faintest pairs of points, for which observational biases could modify 
the correlation, we find 
that the correlation coefficient between pairs of measurements of the same object is
0.94. The slope of the relation is consistent with unity within 2$\sigma$ (0.93$\pm$0.03).
The mean error in the comparison among different observations of the same object is 0.28 mag.
We conclude that the FAUST measurements in the Virgo cluster area are internally consistent 
over at least a six-magnitude range.
 
We compared our detections with those listed by Bowyer \etal (1995) in the FAUST
catalog. We find that the absolute majority of our sources have been listed in the
catalog, but 8\% (16) were detected here while having no catalog counterpart. As
mentioned above, we used here a different automatic detection criterion than 
used by Bowyer \etal In most cases, these sources are faint, or appear in a single
FAUST image, or both; they may be artifacts but we accept them here as genuine 
as there is no clear reason to reject them.
Table 1 lists in its last column
the FAUST catalog number(s) of sources corresponding to our detections. Again, in most
cases the proposed identifications coincide. 

We found that a number of sources listed 
in the FAUST catalog of Bowyer \etal (1995) could be ``merged'' as a single source.
These presumably are the result of the automated processing of FAUST catalog sources 
and in all cases are sources which appear in more than a single FAUST image.
An extreme example is V66=N4438, which appears on three FAUST images and
was listed in Bowyer \etal as three separate sources.


The identification of the detected sources is based mainly on positional 
coincidence with catalogued objects. We used 
various catalogs of stars, such as 
the Smithsonian Astrophysical Observatory's (SAO 1966),
the Hipparcos Input Catalog (Turon \etal 1993), the TD-1 catalog
(Thompson \etal 1978)),
{\it etc.} We identified
galaxies using the 
Third Revised Catalog of Galaxies (de Vaucouleurs \etal 1991; RC3)
 and the Principal Galaxy Catalog (PGC; Paturel \etal 1992),
mainly through the LEDA data bank. In addition to positional coincidence (3' error box), 
we applied some astrophysical constraints. For example, if near the position of the UV source the
catalogs listed a bright late-type
star, but slightly further away we found a UV-bright object such as a sub-dwarf, we 
accepted the latter as the proper counterpart.

In cases when no suitable catalogued counterpart could be found
within 3' of the FAUST position,
we inspected the Palomar Sky
Survey (PSS) prints at the Wise Observatory. We searched in the immediate neighborhood
of a UV source location for (a) relatively bright objects, with (b) bluish colors, as 
could be determined from a comparison of the E and O plates. These candidate
counterparts were observed at the Wise Observatory 
(see section 3, below).

\subsection{Comparison of FAUST and GUV measurements}

Among the various UV observations toward the Virgo cluster, those by GUV (Kodaira 
\etal 1990) have a very similar bandpass and may be used to check external consistency, 
along with those objects observed by IUE.  The GUV experiment operated during 
two observation periods. The 
"acquisition" (A) phase and the "pointing" (P) phase. During the A phase, 
with the payload scanning and being spin-stabilized, 39
sources  and one upper limit were detected  and identified with
 known stars (Table 1, Onaka \etal 1989).
The FAUST experiment detected 8 of these stars; the remaining
GUV sources lie outside the field observed by FAUST. In the pointed phase
the telescopes were kept within one degree of their intended pointings, with a drift rate of 
0'.5 s$^{-1}$. The final resolution of the GUV P phase images, after stacking
70 images with 2.52 sec exposure each, was $\Delta\alpha$=16'.0 and 
$\Delta\delta$=8'.2.
 
Among the sources detected during the pointing phase and listed in Table 2 of
Onaka \etal (1989), five were detected by FAUST and identified with
stars. An improved reduction of the P phase data led to the
identification of some 42 galaxies and of further anonymous sources
(Table 1 and Table 4  of Kodaira \etal 1990).
FAUST detected 23 of the galaxies of Kodaira \etal (1990) and four of the 
anonymous sources have been identified with known stars. Thus the total number 
of detected sources in common by the two experiments is 39 and one upper limit.

Nine galaxies detected by GUV were  missed by FAUST: 
IC3255, N4360, IC3371, N4458, N4459, N4473, IC 3425, IC3457, N4503, IC3483, IC3501.
IC3255 is listed as FAUST source 3110 by Bowyer \etal (1995) with a S/N$<$4.4. 
For this reason  it had not been selected by the algorithm which produced 
the present list.  The remaining sources do not appear in the  Bowyer \etal (1995)
catalogue.  As for the anonymous objects of Table 4 of Kodaira \etal (1991), five   
(A1221+09.7, A1221+12.7, A1224+14.2, A1231+10.7, 
A1235+10.2) were missed here as well as in the FAUST catalog (Bowyer 
\etal 1995). The P10 source
(Table 2 of Onaka \etal 1989) is listed as catalog source 3195,
but is missed in the present list, due to low S/N.

Below we comment on several sources, for which the better
resolution of FAUST allows an improved identification.  The two GUV sources 
identified as N4387 and N4402, and the upper limit 
source identified as N4406, overlap. 
FAUST detected only one source, V51, better identified as N4406.
 The GUV source identified as N4435 is interacting with N4438, which is 
 tidally disturbed and is FAUST V66. 
  IC3416 is 5' from V74, which is identified here as a F0 star.
 N4497 is not a good identification. The FAUST source V90 is identified 
with IC3446, which is nearer the UV location than N4497.
  N4564 was identified as the GUV source, but it overlaps with  N4567 and N4568. 
N4567 is FAUST V126.  IC3583 overlaps with N4569=V129.  The GUV source A1224+08.8 
lies in the field of V56 and V62. See the comment 
in section 3.1. The source A1225+11.5 (P2) is 4' south of V64 and can be identified
with HD108452 (A0). Note that Fig. 5 of Onaka \etal (1989) is reversed and North is
at the bottom. A1225+11.8 (P3) is close to V65. A1227+08.6 (P7) corresponds to V82. 
The anonymous object suggested by Onaka \etal
(1989) has been observed to be a late-B star [S122710+083901].
A1230+13.2 is V102, observed to be a late-B star [S123030+131240].
 A1232+09.7 (P17) corresponds to V122. The object near the UV
position has been observed to be a late-B star [S123252+094327].
 
Our analysis shows that, unfortunately, GUV cannot serve as an external confirmatory source
for FAUST. This is a result of the FAUST exposures being deeper and having better resolution,
as well as being calibrated against IUE measurements. Despite
those aspects, it is reassuring to see that (a) most ``real'' sources of GUV {\bf were}
found by us, and (b) spurious GUV sources, with low S/N, {\bf were not} detected by two different
source-finding algorithms in the FAUST images; that of Bowyer \etal (1995) and the one used by us.

\section{Optical observations}



One hundred and forty six sources were identified with stars or galaxies, as explained
above. The remaining 45 FAUST detections were inspected on both PSS copies at the Wise
Observatory, to identify possible counterparts below the magnitude limit of the catalogs
we consulted. Spectroscopic observations of these candidates from the
Wise Observatory resulted in 31 sources identified as stars, which we
classified as hot stars. In 7 FAUST fields we did not observe all
possible stellar candidates (V1, V65, V92, V99, V112, V119, and V180).
 Two other FAUST sources were found
to have no reliable optical counterpart on pure astrophysical reasons: these are
V101 and V154, which are discussed in sub-section 3.1 below. The remaining
5 sources (V56, V62, V104, V116, and V125) are discussed below is section 3.1. All
these 14 sources are labelled "NO ID" in Table 1.

We used the FOSC (Faint Object Spectrometer Camera)
in spectroscopic mode for these observations. The instrument
 consists of a focal reducer camera with a collimated 
beam section and long-slit focal plane entrance apertures. 
The FOSC operates at the f/7 focal position of the 1 meter telescope
of the Wise Observatory.  It is possible to insert 
grisms in the collimated beam section to disperse the incoming light.  
We used  grisms with 300 gr/mm or 600 gr/mm, and a wedged window as beam steerer 
to center the desired spectral region onto a TI CCD chip. The CCD has a 1024x1000 pixel 
format, but for spectral observations only the relevant part of the chip, containing 
the spectrum of the object and that of the nearby sky, was read and recorded. The FOSC 
spectra in the first configuration cover the region from $\sim$4000\AA\, to $\sim$8000\AA\, 
with $\sim$4\AA\, per pixel. Subsequently, the TI CCD was changed to a Tektronix 
1024$\times$1024 pixel back-illuminated CCD with a Metachrome II coating to enhance its blue-UV 
response. The pixels of this chip are larger than those of the TI. In this configuration 
the spectra cover the range from $\sim$3750\AA\, to $\sim$7100\AA\, with 
$\sim$3.5\AA\,/pixel.

Each object was observed in a slitless configuration (thus the seeing disk defined the
 entrance aperture). We observed spectrophotometric standard stars and obtained calibration 
exposures of a He--Ar arc with a 2" slit together with the candidate sources. Flat fields 
were obtained of the twilight sky and of an internal incandescent filament lamp. The spectra 
were extracted from the CCD images after debiasing and were flat fielded and divided 
pixel-by-pixel by a similarly processed spectrum of the standard star. Figures 3a to 3d 
show representative spectra obtained for some survey sources. 
In the spectrum of V24 it is possible to see a flux depression near 4650\AA\,.
 This is an artefact caused by the standard
star used to reduce the observations of that particular night. The flux depression
appears in all spectra obtained during that specific night.

Most spectra show only Balmer absorption features. The lines are pronounced, but {\bf not} 
extremely wide, indicating that the objects are probably not DA white dwarfs. No HeII features are 
observed. We compared the FOSC spectra with ``template'' spectra from Jacoby \etal (1984) and 
from Silva and Cornell (1992). 
Most appear to be early--type stars, from mid--B to late--A or F. The resolution is not 
sufficient to distinguish 
between main sequence and giant luminosity classes, but supergiants appear to be ruled out
in most cases, because supergiant B stars 
have almost no H$\alpha$ absorption while late A supergiants show reasonably deep HeI 
$\lambda$5876\AA\, absorption. The WiseObs spectral types of all the candidate
optical  counterparts are listed 
in Table 2, along with those typing data obtained from the literature.

The distribution of all FAUST UV  sources in the Virgo fields, detected here and identified as
described above, is summarized in Table 3. We bin the sources by spectral type and UV
magnitude (if they are stars), or just by magnitude if they are extragalactic. We also log
the 14 non-identified sources.

\subsection{Individual objects}

Here we discuss selected individual objects that are worth some extra entry that could not 
be fitted in the tables.

\begin{itemize}

\item V44=N4374=M84 is an elliptical galaxy with a skewed, extended dark lane in its central 
region. The dark lane is associated with a rotating gas disk with a LINER spectrum. The extinction 
by dust in the
dark lane was studied by Goudfrooij \etal (1994), who concluded that the dust was similar with 
that in the Galaxy. The total amount of dust was estimated at $\sim$3 10$^4$ M$_{\odot}$. The 
galaxy also contains significant amounts of hot gas, being an EINSTEIN X-ray source (Forman \etal 1985;
Matilsky \etal 1985). Despite its dark lane, its outer isophotes are remarkably uniform (Malin 1994).

\item V46=N4383=U7505=Mrk 769 is a bright Seyfert galaxy with a starlike, blue nucleus (Barbieri
and Benvenuti 1974). A comparison of the [FAUST--B] and [UV--B], using the Smith and Cornett (1982)
magnitude, indicates a very blue spectral energy distribution (SED). 

\item In the field of V56 and V62  there are some possible candidate counterparts of UV sources.
The faint galaxy between the two sources has been observed spectroscopically.
We found a single emission line at  $\lambda$5960\AA\,. One of the two stars is  SAO119413 (K0) and
the other, S122425+085025, has been observed spectroscopically once from the WiseObs
and was identified as a late A star.
Kodaira \etal (1990) found a UV source which they did not identify (A1224+08.8) about
3' South of V56; it is possible that this GUV source and V56 are the same, and coincide 
with the late-A star.

\item V65: The two m$\approx$13 mag candidates were not observed. There is one
dwarf galaxy (VCC 1020; dE) which is nearby. If it is adopted as a possible
conterpart, its [FAUST--B] color (--5.67) would be anomalous for a dwarf elliptical.

\item V74: The observed candidate S122644+105941 is a V$\simeq$12.5 late F star. In
the field there is a faint blue star near the UV source position, and two other 
candidates 6 arcmin farther away. The faint V$>$15 blue candidate, S122629+105705, has
 not been observed.

\item V79: The star S122700+063417 has been observed and it is a late F
 star. The suggested identification is near the UV position, the V=8.7 mag
 G0 star SAO119429.

\item  V92: The S122903+080353 object near the UV  position has been observed spectroscopically
and identified as a $\sim$10 mag late G star. There are three other candidates in the field, which were
not observed. One of these, fainter than V=15, appears extremely blue when
compared on the PSS E and O copies.

\item V99: On the PSS there is a faint blue and slightly compact galaxy
 about 1'.4 E of the UV position,  which was not observed.

\item V101: The faint object S123017+094747, which is near the UV position, has been
 observed and identified as a late G star, fainter than V=15 mag.

\item  V104: The S123031+051936 object near position has been observed to be 
a V$\simeq$13.5 mag F/G star. There are other fainter candidates in the field, 
which were not observed.

\item V106 was identified as the QSO Q1230+0947, a V=16.2, z=0.420 AGN found in the APM
objective-prism survey (Foltz \etal 1987).

\item V116: The V=14 star S123153+154409 has been observed and it is an
early G star. There are some other faint candidates in the field.

\item V119: A m$>$15 blue star is located $\sim$1' away from the UV position, 
and was not observed.

\item V121: A very weak spectrum was obtained from the m$>$15 mag candidate in 5400 sec,
which could not be used for classification. A dwarf elliptical galaxy (VCC 1600) is
nearby and if selected as counterpart would have a very anomalous color [FAUST--B]=--5.68
for a dwarf elliptical, leaving the association of V121 with VCC 1600 uncertain.

\item V125: The S123404+075544 candidate observed by us is a V$\simeq$13 mag late-G star. This 
is the brightest candidate object among the star-like images on the PSS near the location of the
UV source and it has a faint bluish companion to the West-South West.

\item V129=NGC 4569=M90=UGC 7786 is an Sab galaxy. It was observed during a snapshot FOC-HST program 
(Maoz \etal 1995) to have a nuclear UV source at 2300\AA\, whose size is 
FWHM$\leq$0".22=2.2$h^{-1}$ pc
({\it h} is the Hubble constant in units of 100 km s$^{-1}$).
Spectroscopy of the source indicates it to be a transition between an HII nucleus and a LINER. There is
residual UV emission from an extended source 0".65 South of the bright nuclear region. The flux density
measured for the nuclear source is 10$^{-14}$ erg/sec/cm$^2$/\AA\,, or a monochromatic magnitude
at 2270\AA\, of 13.82. The difference from the integrated UV flux measured by FAUST is due to
emission at much larger nucleocentric distances than sampled by the FOC. 

\item V138=NGC4579=M58=UGC 7796 is an Sab galaxy with a LINER nuclear spectrum and evidence of weak broad
 H$\alpha$ and X-ray emission (Fillipenko and Sargent 1985, Halpern and Steiner 1983). It was observed 
in the same FOC-HST snapshot survey to have two near-nuclear sources, a bright nucleus and a nearby
slightly extended source (Maoz \etal 1995). The FOC nuclear source flux density corresponds to a
monochromatic magnitude at 2270\AA\, of 16.22 mag. The slightly extended source is two magnitudes fainter.
However, the FAUST measurement indicates that most UV emission does not originate from the
nuclear region.

\item V144=VCC 1791=UGC 7822=IC3617=MCG 01-32-127=DDO 140 has strong H$\alpha$ emission 
and equivalent width, indicating vigorous
star formation activity (Almoznino 1995). We have mapped it with the VLA and found
an HI companion in contact with the main galaxy (Brosch \etal 1996). 

\item V152: A very weak spectrum was obtained in a one hour exposure for the observed candidate
(S123754+101551), a V$>$15 bluish object which coincides with the location of the UV source.
The color [FAUST--V] is very blue for the assigned spectral type; this may indicate some
source peculiarity.

\item V154: The S123837+071338 candidate observed is
a V$\simeq$13 mag K star. No other candidates are present in the field. The UV flux appears 
to be fairly well established, however note that the source appears on a single FAUST image.
In view of the faintness in the visible, {\it i.e.,} the very blue [FAUST--V] color of 
this K-star, we adopt a ``no identification'' label for the source.

\item V169: A 5400 sec spectrum showed Balmer lines and a steep continuum, as expected from a 
B type star. The object is very faint (V$>$15 mag) and, if selected as counterpart, would imply
 [FAUST--B]$<$--1.75.

\item V171=N4649+N4647=Arp 116 is a contact pair of galaxies consisting of the
S0$_1$(2) N4649=M60 and the Sc  N4647.
The S0 galaxy is included in the list of objects for which Storchi-Bergmann, Kinney and Chalis (1994)
produced the UV to near-IR spectral distributions. This is based on a combination of 
IUE and ground-based spectrophotometry using an entrance aperture shaped and oriented to match that 
of IUE's.  Using their flux density at 1583\AA\, (the spectral region closest to FAUST's)
we obtain a monochromatic magnitude of 15.03, three magnitudes fainter than measured by FAUST but more
than one magnitude brighter than the FOC HST snapshot measurement (Maoz, 
private communication). Part of the difference could be attributed to 
the shape of the spectrum in the UV, but note that the spectral slope as shown by Storchi-Bergmann
\etal stays reasonably flat shortwards of 3000\AA\,. This implies that the UV emission is this LINER is
extended, with a small fraction emitted in the nuclear region sampled by the FOC or IUE, 
and the entire emission detected by FAUST.
Our FAUST total magnitude is consistent with measurements, albeit at different UV
bands, from SCAP (m$_{UV}$=12.0) and Smith and Cornett (1982) (m$_{UV}$=11.87). 

\item V172=NGC4651 is an Sc galaxy that, at v$_{\odot}$=800 km s$^{-1}$, is probably in the
VC. Deep plates ({\it e.g.,} Sandage \etal 1985) show the presence of a jet-like feature. Malin 
(1994) found a faint outer shell at the end of this structure. Giuricin \etal (1990) include 
V172 in their LINER list, while Kenney and Young (1988) note that it has a high SFR.

\item V180: A 3600 sec spectrum of a faint bluish candidate identified on the
Palomar prints shows only a weak continuum with no emission lines. No spectral
classification is possible from this spectrum.

\end{itemize}

\section{Discussion}

\subsection{Stars in the direction of the Virgo cluster}

Table 3 gives the distribution of 
stars according to spectral type and magnitude. In comparison with our previous FAUST study
of the North Galactic Pole region (NGP: Brosch \etal 1995), one item is immediately apparent: no white 
dwarfs were positively identified in the VC fields. The NGP FAUST field covered 69 
square degrees; we detected there 81 objects, thus 1.17 per square degree ($\Box^{\circ}$). 
The source density
in the $\sim100\Box^{\circ}$ VC fields is higher, 1.91 per $\Box^{\circ}$. However, the VC images are toward a 
cluster of galaxies, whereas the NGP image is of the general field. Excluding galaxies 
from both fields, and
concentrating only on identified objects, we find a UV star density of 1.03 stars per $\Box^{\circ}$
at the NGP {\it vs.} 0.96 stars per $\Box^{\circ}$ toward the VC.  

The density of detected and identified FAUST stellar UV sources in the direction of 
the VC does not differ 
from that to the NGP. Despite this, no white dwarfs (WDs) are identified in the VC UV images. The
space distribution of (DA) WDs was studied by Boyle (1989) and was included in the model
of the UV sky (Brosch 1991). We modeled the stellar distribution for the VC central location
using the FAUST spectral bandpass and show a comparison of the detected {\it vs.} the 
predicted stellar density in Figure 4.
The model follows the actual stellar distribution down to m$_{UV}\simeq$11,
where incompleteness sets in, but does not reproduce it with great fidelity (half the data
points up to the completeness limit are more than 1$\sigma$ from the model). It is
possible that this is a ``small-numbers'' effect because of the size of the sample,
since
according to our model, 
there should be at least three WDs in a 100$\Box^{\circ}$ wide field 
up to a UV magnitude limit of 14. About 140 stars of all types,
to this magnitude limit, should have been detected. Considering the completion 
limit of FAUST, the model and
the total number of detected stars agree. The discrepancy with the expected number of WDs remains unsolved.
One possibility is that the missing WDs are among the NO ID sources, another is that some were 
mis-classified by us as A or B stars, where in reality they were DAs ({\it e.g.}, V152, V169 or V186).
Another is that WDs are observationally discriminated against,
 when searched for in the vicinity of a cluster 
of galaxies. A similar ``avoidance'' of Galactic stars by QSOs (in objective-prism surveys)
was demonstrated by Gould, Bahcall and Maoz (1993).

V24=S121757+101353 has been shown by our spectroscopic observations to be an early-G V=13.2 object.
Its spectrum is shown in Fig. 3a. Note that the 4650\AA\, absorption-like
feature is an artefact (see section 3).
The [FAUST--V] color, --5.72, is very blue for such a spectral type, where more yellowish colors
are expected. The source was detected on two separate FAUST images and its count statistics is
reasonable. We have therefore reasonable confidence of its reality and of the accuracy of 
the UV magnitude. The visual magnitude is from the HST Guide Star Catalog; its accuracy is of 
order 0.1--0.2 mag (Ratnatunga 1990). Such a blue UV--V color index is expected only from a
``mixed'' source, where a late-type optically-bright star is paired with a hot companion, which
outshines the late-type star in the UV. This star is, therefore, a candidate for such a mixed binary
object. Note that our spectrum does not show a trace of such a hot companion; spectra reaching
further into the blue-violet are required to detect this.

V143=SAO100195 is an A0m star, for which we found [FAUST--V]=4.42. Am stars are those ``A or early
F stars for which no unique spectral type can be assigned'' (Jaschek and Jaschek 1987).
The color we measured is much redder than expected for ``regular'' A0 stars, which have 
colors near 0.0; in absence of 
interstellar reddening, which is not expected at the galactic latitude of this star, the 
possibilities could be either circumstellar reddening or some photospheric activity 
that reddens the UV--optical colors. However, Table 2 shows that three other Am stars (V3, V120 
and V178) also have positive and fairly large [FAUST--V] color indices. 
It is possible that this is caused by 
additional metallic lines present in the UV spectrum of Am stars; this has been noted by Jascheck
and Jaschek (1987) for the optical spectra and is probably true also in the UV.

\subsection{Virgo galaxies: UV and other observational properties}

In Table 4 we present the UV properties of the galaxies detected by FAUST with morphological
types from the BST catalog. We 
include UV measurements from the GUV experiment (Kodaira \etal 1990), from the SCAP
balloon telescope (Donas \etal 1987), and from Goddard rocket-borne 
telescope (Smith and Cornett 1982, labelled S+C). For consistency, we transformed all 
GUV  measurements into monochromatic magnitudes as explained above.
Note that each experiment
measured a different spectral segment; GUV measured effectively at 1560\AA\, with a 
bandwith of FWHM=230\AA\,, SCAP measured at $\sim$2000\AA\, and the bandpass was 
$\sim$200\AA\,, and the Goddard imager had a 1150\AA\, wide bandpass centered near 
2421\AA\,. 

We mentioned above that, in a number of cases, the UV emission detected by FAUST is
extended, and from a comparison with HST FOC observations only a small fraction of the UV
is centrally produced. This was shown to be the case for a number of elliptical
galaxies in the VC by Kodaira \etal (1990), who compared the GUV flux density with that measured 
by the large IUE aperture. They found that, for five galaxies, the ratio of central to total UV flux
ranged from 14\% to more than 63\%.  This is confirmed here by a comparison of FOC 
(Maoz \etal 1995)
and other UV observations for three of the galaxies, as mentioned above.
Observations by UIT during the ASTRO I and ASTRO II missions (e.g., Neff \etal 1994,
Cornett \etal 1994) detected complex morphologies in the far-UV
emission of some galaxies. In the case of M31 and M33, far-UV
observations found also that their central regions were
bluer than their outer parts (O'Connell 1992).

The galactic latitude of all sources discussed here is higher than 65$^{\circ}$.
Despite this, the UV measurements could be affected by foreground
galactic extinction. Burstein and Heiles (1982) showed that small 
but significant reddening may exist over the Virgo cluster. We 
consulted the reddening catalog of Burstein and Heiles (1984)
and found that
our objects have small reddening values
[E(B-V)$<$0.04]. This corresponds to A$_{1650\AA\,}\leq$0.33 mag,
using the wavelength dependence of the extinction from Savage and
Mathis (1979), {\it i.e.,} $\frac{A(1650)}{E(B-V)}=8.0$.
 We corrected the FAUST magnitudes for extinction, using the 4$\times$E(B--V) 
values listed for the galaxies by Burstein
and Heiles, or as appropriate for the nearmost galaxy listed by them in 
cases where no specific entry appeared in the list of Burstein and
Heiles. We corrected by A$_{UV}$=2$\times$ the listed value if positive, or
left the UV magnitude unchanged if Burstein and Heiles listed a zero or
negative value. The UV magnitudes, still called ''FAUST magnitudes'' and
 corrected for extinction, are listed in 
Table 5, where galaxies are discussed separately from Galactic
objects. The table also presents the total magnitude of the galaxy
(usually in the B band and from the BST or the RC3 catalogue), corrected
for extinction, and
adds the T-type of each and 
the HI flux integral FI(HI), from Huchtmeier and Richter (1986), Haynes and Giovanelli (1986),
 Hoffman \etal (1989a and 1989b), or Hoffman (private communication). In the   
case whenever a number of FI(HI) values are given we adopted their straight average. The correlation 
between FI(HI) and the FAUST UV magnitude is shown in Figure 5. We find a correlation coefficient
of --0.74 when regressing $\log$ FI(HI) against the FAUST magnitude, and including only
spiral and Irregular galaxies (T$>$0). This has been known from other UV--HI studies (\eg \,
Buat \etal 1989) and has been linked to a correlation between total gas content and the star
formation rate (see below).

In Table 5 we include also the far infrared (FIR) measurements of the same galaxies by
the IRAS telescope. The data are mainly from the IRAS Faint Source
Catalog. We estimate the total FIR flux as in Lonsdale \etal (1985):
\begin{equation}
F_{FIR}=1.26 \, (2.58S_{60}+S_{100}) \, 10^{-14} \, W \, m^{-2}
\end{equation}
using the 60 and 100 $\mu$m flux densities, which are also given in Table 5. The correlation
between the FAUST and the FIR emissions is shown in Figure 6. 
We plot in Figure 7 the distance-independent ``colors'' $\frac{FIR}{HI}$ {\it vs.} [FAUST--B].
Excluding three background galaxies at high {\it cz} (V12, V63, and V118), 
which may be exceptional, we find a correlation coefficient of 0.62. 

The color [FAUST--B] can be used, as done by Deharveng \etal (1994), to compare the VC galaxies 
with the stellar population models of galaxies which are used for deriving the K-correction
in cosmological models. Deharveng \etal compared the colors with the spectral energy distributions 
of Coleman, Wu and Weedman (1980).  
Our study, like that by Deharveng \etal, and previously by Faber (1982) and by
Burstein \etal (1988), shows that elliptical galaxies exhibit a large dispersion
in their (UV-B) color index relative to other, late-type galaxies. Thus, it is hard to establish 
a universal K-correction for their UV segment of the spectrum, which could be useful when studying
in the optical elliptical galaxies at z$\simeq$2. Hence, the elliptical galaxies do not form a
uniformly bright population of objects at high redshifts, making their use in optical surveys or
studies of high redshifts objects problematic, due to Malmquist and other selection effects.

\subsection{Total UV flux from cluster galaxies}

In the background section we explained 
that the discovery of nearby Lyman $\alpha$ absorptions in the spectra of  
3C273 prompted the evaluation of the metagalactic ionizing background. The
presence of such absorbers in the direction of the VC and at the redshift of the cluster
(Weymann \etal 1995, Shull \etal 1996)
prompts us to re-evaluate the ionizing flux, based on the measured UV emission from the
cluster galaxies.

Here we restrict the discussion to the total UV flux detected from  galaxies
in the direction of the VC by different UV experiments.
We estimate the total 
number of available ionizing photons, based on published UV data  
of Virgo cluster galaxies. As the bandpasses of observations were always {\it longward} of the Lyman 
limit, we estimate the Lyman continuum (Lyc) photon flux by scaling the measured fluxes to the 
wavelength of the Lyman break using the stellar population models of evolving galaxies of Bruzual 
and Charlot (1993). 

Each publication with UV observations of the VC 
provides a list of galaxies towards the Virgo cluster and either 
a UV ``monochromatic'' magnitude or a flux density in a specific bandpass. In order to estimate 
the amount of diffuse UV contribution when the 
locations of the galaxies are not known in three-dimensional space, we 
{\it assume} that the  UV contribution from all
 galaxies comes from a single point located at the cluster center. We calculate
the UV contribution from each galaxy from the given flux, we scale it for the distance
of each galaxy to the center of the cluster (to account statistically for nearer
and farther galaxies than the center of Virgo),
and add it to the total flux densities (TFD).
The reference position for the center of the Virgo cluster was given by van den Bergh (1977):
$\alpha$=12$^h$ 25$^m$.4, $\delta$=+12$^{\circ}$ 40'.
Objects for which only an upper limit is 
given are included as if their emission would equal this upper limit (except for FAUST). 
This yields an upper limit 
to the total galactic UV emission of the Virgo cluster. 

The TFD reached for the different instruments/wavebands
are remarkably similar; for the detected FAUST galaxies the TFD is 4.82 10$^{-12}$ erg sec$^{-1}$ 
cm$^{-2}$ \AA\,$^{-1}$, whereas for the objects listed by Smith and Cornett (1982) the TFD is 
$\sim$2 10$^{-12}$ erg sec$^{-1}$ cm$^{-2}$ \AA\,$^{-1}$. The Virgo galaxies 
measured by SCAP and listed in Donas 
\etal (1987) total to a TFD of 1.1 10$^{-12}$ erg sec$^{-1}$ cm$^{-2}$ \AA\,$^{-1}$, but only 
a few brighter objects are included in their list. The TFD for the GUV galaxies is $\sim10^{-12}$ 
erg sec$^{-1}$ cm$^{-2}$ \AA\,$^{-1}$, close to the SCAP 2000 total.

The TFD measured by FAUST is $\sim2.5\times$ larger than that estimated by Weymann \etal (1995)
for the 1550\AA\, flux of VC galaxies. Weymann \etal calculated the flux from 
the B-band luminosity of early
type galaxies, compounded with a UV--V transformation derived from Burstein \etal (1988).
The difference from the FAUST measurement is the result of the neglect of spiral and 
other late types by Weynmann \etal;
these galaxies produce most of the UV flux of the VC in the FAUST band.

We adopt the highest value of TFD from FAUST, and scale it to the Lyman limit, using
 the maximal ratio of 1650\AA\, flux to Lyc flux just short 
of the Lyman limit allowed by the galaxy models of Bruzual and Charlot (1993), 
which is $\sim$2. From this, we conclude that the TFD at the Lyman limit 
which is produced by galaxies detected at Earth can be at most $\sim$8 
10$^{-12}$ erg sec$^{-1}$ cm$^{-2}$ \AA\,$^{-1}$, the equivalent of
a single m$_{UV}$=7.2 mag star.  

\subsection{Dust in the Virgo cluster ?}

Girardi \etal (1992) found, by analyzing loose groups, an anticorrelation between
the E(B-V) color excess of galaxies and the velocity difference with respect to the
brightest group member. They interpreted this as evidence for the presence of 
dust within the intra-group space, which reddens the light from galaxies in the
anti-Earth location. Similar claims, of significant amounts of dust within Abell clusters,
were made by Karachentsev and Lipovetski (1969), Bogart and Wagoner (1973), Boyle, Fong and 
Shanks (1988), and Romani and Maoz (1992). In the context of the VC, a search for inter-cluster
dust can be linked to the possible presence of ``almost invisible'' galaxies, which
give rise to the absorption lines at the redshift of the VC in spectra of background QSOs.

We attempted to test this by first checking five background galaxies, with v$_{\odot}\geq$3000 km s$^{-1}$,
against cluster and foreground galaxies with v$_{\odot}<$3000 km s$^{-1}$, hereafter called
``comparison galaxies'' (CGs). The assumption here is that background galaxies, seen through the
cluster, should have redder [FAUST--B] colors that the CGs of similar morphological types
(within 2 T subclasses), because of dust reddening. We found that
there is no consistent evidence for the background galaxies being significantly redder than the
cluster and foreground galaxies.

We then tested the CGs against themselves, by splitting the sample into a ``near'' bin if
v$_{\odot}>$1100 km s$^{-1}$ (redshift of the Virgo core), but v$_{\odot}<$3000 km s$^{-1}$,
and a ``far'' bin, with galaxies which have v$_{\odot}\leq$1100 km s$^{-1}$. The assumption here is 
that galaxies which belong to the cluster have their cosmological recession 
velocity modulated by the gravitational
attraction of the cluster (see, \eg \, Binggeli \etal 1993). 
Therefore, galaxies that are between us and the VC core will be 
accelerated by the cluster attraction and will have, on average, velocities higher than 
that of the cluster core. Galaxies in the VC but beyond the core will be slowed by the 
cluster attraction and may even have negative heliocentric velocities.
Although the gravitational influence of the cluster is certainly much more complex,
this crude separation may show some significant result. The separation, and subsequent 
comparison of cluster members, compensates for such effects as ``anaemic'' spirals, 
stripped galaxies, and such, which should apply equally to both the ``far'' 
and the ``near'' sub-samples. In addition, Malmquist biases in the selection of background
objects should be minimized.

We tested whether the ``near'' bin galaxies are bluer in [FAUST--B] than
the ``far'' bin galaxies, but did not find significant differences.
The conclusion is that the total amount of dust in the VC is 
rather small, probably negligible. A similar result for a sample of clusters
was obtained recently by Maoz (1995), from a study of the (V--I) color distribution 
of QSOs beyond the clusters.

\subsection{Large-scale distribution and UV luminosity function of UV-emitting galaxies}

In the context of the three-dimensional distribution of galaxies in the direction of the
VC it is interesting to test the distribution of UV-emitting objects. One finding, emphasized 
here as well as by Deharveng \etal (1994), is that UV-emitting galaxies are mainly of 
late-type. These galaxies show usually less clustering than early-type objects, such as 
elliptical, lenticular and Sa galaxies. It is therefore to be expected that UV-bright
objects will tend to line at the edges of the main clusterings, and in the voids.

In order to check the three-dimensional distribution of UV galaxies we
plot in Figure 8 a diagram of all detected objects. The vertical axis is the 
heliocentric velocity as listed by BST and the horizontal axis is the right ascension 
of each galaxy. It is clear that most of the galaxies detected by FAUST are within
the redshift range of the VC and only a few are in its background.

In addition, we tested the distribution of the cluster galaxies in Virgocentric projected 
distance (degrees from the adopted cluster center position), by binning the galaxies in one-degree wide
rings, despite the fact that our observations cover an $\sim$elliptical shaped sky area.
Considering that galaxies are actually distributed in a three-dimensional volume, 
the distribution indicates a high over-density of UV emitting galaxies in the inner four
degrees of the VC.


The optical luminosity function (LF) of galaxies in the VC has been studied by
Sandage, Binggeli and Tammann (1985). The LF for bright galaxies is well 
represented by a function bounded at both bright and faint ends. The optical LF for field
galaxies was studied by Marzke \etal (1994) and was fitted by single Schechter functions 
(Schechter 1976) over the
entire magnitude range. In the Coma cluster, Biviano \etal (1996) found that the best 
fit is by a Schechter function combined with an Erlang function.

We plot in Figure 9 the distribution of UV magnitudes for cluster
 member galaxies, \ie those with v$_{\odot}<3000$ km s$^{-1}$, 
excluding V121 and V155 (uncertain identifications), after binning it 
into half magnitude-wide bins. In view of the small number of galaxies with UV magnitudes,
we  include all our objects, instead of splitting them into morphological
subclasses.

We note that, for the VC's bright galaxies, Sandage, Binggeli and Tammann
found that a Gaussian best represents the luminosity function. Only when including the
lower luminosity objects, the dwarfs, were they justified in fitting a Schechter function.
The Gaussians fitted by Sandage \etal peak at B$^*$=13.3 and 13.0 for spirals and E+S0 galaxies,
respectively. The dwarf galaxies were fitted by Sandage \etal as Schechter functions,
with $\alpha$=--1.35 and B$^*$=14.3 (for dEs). Studies of the LF of field galaxies (\eg Efstathiou 
\etal 1988, Loveday \etal 1992) found less steep $\alpha$ values (--1.07 and --1.11, respectively).

We fitted a Gaussian distribution to the histogram of FAUST galaxies of the VC, and allowed 
the peak, the average, and the width of the Gaussian to change. A best fit was estimated by 
minimizing $\chi^2$, where the average UV$^*$=12.37 and the full width at half-maximum of
the Gaussian was 2.02 mag.  
 
\section{Conclusions}

We presented a comprehensive study of three UV FAUST images in the direction of the Virgo
cluster of galaxies. We detected 191 UV sources and identified them, through comparison
with catalogued optical sources, with (mostly) early-type stars or galaxies. We showed that,
within the completeness limit of FAUST observations, the UV stellar distribution follows that
predicted by our UV sky model. Am stars were shown to be UV--faint, in comparison
to normal stars of the same spectral sub-class.

For galaxies, we confirmed that significant UV flux is emitted outside the central regions.
We confirmed earlier results that elliptical galaxies have widely varying UV brightness,
making their use in studies of high redshift galaxy samples problematic.
Star formation properties were demonstrated through UV--IR and UV--HI correlations. These
confirm essentially the accepted pattern of star formation. We estimated the total ionizing
flux by galaxies in the VC and showed that it is consistent with earlier estimates. We could 
not put significant new limits on dust in the VC 
but our results are consistent with a ``no dust'' case. The UV galaxies were found to
concentrate in the central region of Virgo. The UV luminosity function was fitted with 
a Gaussian. A better determination of the
UV luminosity function in the Virgo cluster requires deeper UV observations.

\section*{Acknowledgements}

The UV astronomy effort at Tel Aviv is supported 
by special grants to develop a space UV astronomy experiment 
(TAUVEX) from the Ministry of Science and the Arts, and from
the Austrian Friends of Tel Aviv University. Observations at the Wise Observatory
are supported by a Center for Excellence Grant from the Israel Academy of Sciences. 
This paper made use of the  LEDA galaxy data bank.
Many optical observations
for this project were mostly performed by Mr. E. Goldberg and the reductions 
by Ms. Ana Heller. Some of the identification,  using 
computerized data banks, were done by Ms. Susanna Steindling. 
NB is grateful to J. Donas for providing a list
of galaxies observed by SCAP, to G. Lyle Hoffman for an updated
list of galaxies in the VC region with HI measurements, and to B. Bilenko for
running the UV sky model for the VC.

\section*{References}
\begin{description}

\item Almoznino, E. 1995 PhD thesis, Tel Aviv University.



\item Barbieri, C. and Benvenuti, P. 1974 Astron. Astrophys. Suppl. {\bf 13}, 269.

\item Binggeli, B., Popescu, C.C. and Tammann, G.A. 1993 Astron. Astrophys. Suppl. {\bf 98}, 275.

\item Binggeli, B., Sandage, A. and Tamman, G.A. 1985 Astron. J. {\bf 90}, 1681 (BST).

\item Binggeli, B., Tammann, G.A. and Sandage, A. 1987 Astron. J. {\bf 94}, 251.

\item Biviano, A. Durret, F., Gerbal, D., Le Fevre, O., Lobo, C., Mazure, A. and Slezak, E.
1996 Astron. Astrophys. {\bf 311}, 95.

\item Bogart, R.S. and Wagoner, R.V. 1973 Astrophys. J. {\bf 181}, 609.


\item Boyle, B. 1989 Mon. Not. R. astr. Soc. {\bf 240}, 533.

\item Boyle, B.J., Fong, R. and Shanks, T. 1988 Mon. Not. R. astr. Soc. {\bf 231}, 897.

\item Bowyer, S., Sasseen, T.P., Lampton, M. and Wu, X. 1993 Astrophys. J. {\bf 415}, 875.

\item Bowyer, S., Sasseen, P.T., Xiauyi, W. and Lampton, M. 1995 Astrophys. J. Suppl.
{\bf 96}, 461.


\item Brosch, N. 1991 Mon. Not. R. astr. Soc. {\bf 250}, 780.

\item Brosch, N., Almoznino, E., Leibowitz, E.M., Netzer, H., Sasseen, T., Bowyer, S., 
Lampton, M. and Wu, X. 1995 Astrophys. J., {\bf 450}, 137.

\item Brosch, N., Almoznino, E. and Hoffman, G.L 1996, submitted.

\item Bruzual, G.A. and Charlot, S. 1993 Astrophys. J. {\bf 405}, 538.

\item Buat, V., Deharveng, J.M. and Donas, J. 1989 Astron. Astrophys. {\bf 223}, 42.

\item Burstein, D., Bertola, F., Buson, L.M., Faber, S.M., and Lauer, T.R.
1988 Astrophys. J. {\bf 328}, 440.

\item Burstein, D. and Heiles, C. 1982 Astron. J. {\bf 87}, 1165.

\item Chakrabarti, S., Sasseen, T.P., Lampton, M. and Bowyer, S. 1993, Geophys. Res. Lett. {\bf 20}, 535.

\item Coleman, G.D., Wu, C.C. and Weedman, D.W. 1980 Astrophys. J. Suppl. {\bf 43}, 393.


\item Cornett, R.H., O'Connell, R.W., Greason, M.R., Offenberg, J.D., Angione, R.J.,
Bohlin, R.C., Cheng, K.P., Roberts, M.S., Smith, A.M., Smith, E.P., Talbert, F.D.
and Stecher, T.P 1994 Astrophys. J. {\bf 426}, 553.


\item de Vaucouleurs, G., de Vaucouleurs, A., Corwin, H.G., Buta, R.J., Paturel, G.
and Fouqu\'{e}, P. 1991 {\it Third Reference catalog of Bright Galaxies}, New
York: Springer Verlag (RC3).


\item Deharveng, J.M., Sasseen, T.P., Buat, V., Bowyer, S., Lampton, M. and Wu, X. 1994, 
Astron. Astrophys. {\bf 289}, 715.

\item Donas, J., Deharveng, J.M., Milliard, B., Laget, M. and Huguenin, D. 1987 Astron. 
Astrophys. {\bf 180}, 12.



\item Efstathiou, G., Ellis, R.S., and Peterson, B.A. 1988 Mon. Not. R. astr. Soc.
{\bf 232}, 431.

\item Elmegreen, B.G., Elmegreen, D.M. and Montenegro, M. 1992 Astrophys. J. Suppl. 
{\bf 79}, 37.

\item Faber, S. 1982 Highlights Astr. {\bf 6}, 165.

\item Fillipenko, A.V. and Sargent, W.L.W.  1985 Astrophys. J. Suppl. {\bf 57}, 503.

\item Foltz, C.B., Chaffee, Jr. F.H., Hewett, P.C., Macalpine, G.M., Turnshek, D.A.,
Weymann, R.J. and Anderson, S.F. 1987 Astron. J. {\bf 94}, 1423.

\item Forman, W., Schwarz, J., Jones, C., Liller, W. and Fabian, A.C. 1979 Astrophys. J. 
Lett. {\bf 234}, L27.

\item Forman, W., Jones, C. and Tucker, W. 1985 Astrophys. J. {\bf 293}, 102.




\item Girardi, M., Mezzetti, M., Giuricin, G. and Mardirossian, F. 1992 Astrophys.
J. {\bf 394}, 442.

\item Giuricin, G., Bertotti, G., Mardirossian, F. and Mezzetti, M. 1990 Mon. Not. R.
astr. Soc. {\bf 247}, 444.

\item Goudfrooij, P., de Jong, T., Hansen, L. and N{\o}gaard-Neilsen, H.U. 
1994, Mon. Not. R. astr. Soc., {\bf 271}, 833.

\item Gould, A., Bahcall, J.N. and Maoz, D. 1993 Astrophys. J. Suppl. {\bf 88}, 53.


\item Halpern, J.P. and Steiner, J.E.  1983 Astrophys. J. {\bf 269}, 37.

\item Haynes, M.P. and Giovanelli, R. 1986 Astrophys. J. {\bf 306}, 466.



\item Hoffman, G.L. 1989 in {\bf Large Scale Structure and Motions in the Universe}
M. Mezzetti \etal (eds.), Dordrecht: Kluwer, p. 365.

\item Hoffman, G.L., Helou, G., Salpeter, E.E. and Lewis, B.M. 1989a, Astrophys. J. 
{\bf 339}, 812.

\item Hoffman, G.L., Lewis, B.M., Helou, G., Salpeter, E.E. and Williams, H.L. 1989b 
Astrophys. J. Suppl. {\bf 69}, 65.

\item Hoffman, G.L., Lewis, B.M. and Salpeter, E.E. 1995 Astrophys. J. {\bf 441}, 28.

\item Huchtmeier, W.K. and Richter, O.-G. 1986 Astron. Astrophys. Suppl. {\bf 64}, 111.

\item Jacoby, G.H., Hunter, D.A. and Christian, C.A. 1984 Astrophys. J. Suppl. {\bf 56}, 257.

\item Jacoby, G.H., Ciardullo, R and Ford, H. 1990 Astrophys. J. {\bf 356}, 332.

\item Jaschek, C. and Jaschek, M. 1987 {\bf The Classification of Stars}, Cambridge 
University Press, p.220.

\item Karachentsev, I.D. and Lipovetski, V.A. 1969 Sov. Astron. {\bf 12}, 909.

\item Kenney, J.D. and Young, J.S. 1988 Astrophys. J. {\bf 326}, 588.

\item Kodaira, K., Watanabe, T., Onaka, T. and Tanaka, W. 1990 Astrophys. J. {\bf 363}, 422.




\item Lampton, M., Sasseen, T.P., Wu, X. and Bowyer, S. 1993, Geophys.Res.Lett. {\bf 20}, 539.

\item Lieu, R., Mittaz, J.P.D., Bowyer, S., Lockman, F.J., Hwang, C.-Y., Schmitt, J.H.M.M.
1996 Astrophys. J. {\bf 458}, 5.


\item Lonsdale, C.J., Helou, G., Good, J.C. and Rice, W. 1985 {\bf Catalog of galaxies and
quasars observed in the IRAS survey}, Jet Propulsion Laboratory, Pasadena, CA.

\item Loveday, J., Peterson, B.A., Efstathiou, G., Maddox, S.J., and Sutherland,
W.J. 1992 Astrophys. J. {\bf 390}, 338.


\item Malin, D. 1994 in {\bf Astronomy from Wide Field Imaging} (H.T. MacGillivray, E.B. 
Thomson, B. Lasker, I.N. Reid, D.F. Malin, R.M. West and H. Lorenz, eds.), Dordrecht: Kluwer
Academic Press, p. 567.

\item Maoz, D. 1995 ApJ, {\bf 455}, L115.

\item Maoz, D., Fillipenko, A.V., Ho, L.C., Rix, H.-W., Bahcall, J.N., Schneider, D.P. and
Duccio Machetto, F. 1995, Astrophys. J. {\bf 440}, 91.

\item Marzke, R.O., Geller, M.J., Huchra, J.P. and Corwin, H.G. Jr. 1994 Astron. J. {\bf 108}, 437.

\item Matilsky, T., Jones, C. and Forman, W. 1985 Astrophys. J. {\bf 291}, 621.





\item Neff, S.G., Fanelli, M.N., Roberts, L.J., O'Connell, R.W., Bohlin, R.,
Roberts, M.S., Smith, A.M. and Stecher, T.P 1994 Astrophys. J. {\bf 430}, 545.


\item O'Connell, R.W. 1992 in {\bf The stellar populations of galaxies} IAU 
Symp. no. 149 (B. Barbuy and R. Renzini, eds.), Dordrecht: Kluwer, p.233.

\item Onaka, T. and Kodaira, K. 1991 Astrophys. J. {\bf 379}, 532.

\item Onaka, T., Tanaka, W., Watanabe, T., Watanabe, J., Yamaguchi, A., Nakagiri, M., 
Kodaira, K., Nakano, M., Sasaki, M and Tsujimura, T. 1989 Astrophys. J. {\bf 342}, 238.

\item Paturel, G., Vauglin, I., Garnier, R., Marthinet, M.C., Petit, C., 
Di Nella, H., Bottinelli, L., Gouguenheim, L. and Durand, N.  1992
{\it LEDA: The Lyon-Meudon Extragalactic Database}, CD-ROM Observatorie de Lyon.


\item Ratnatunga, K.U. 1990 Astron. J. {\bf 100}, 280.

\item Romani, R.W. and Maoz, D. 1992 Astroph. J. {\bf 386}, 36.



\item Sandage, A., Binggeli, B., and Tammann, G. 1985 Astron. J. {\bf 90}, 1759.

\item Sandage, A. and Tammann, G.A. 1974 Astrophys. J. {\bf 194}, 559.

\item Savage, B.D. and Mathis, J.S. 1979 Ann. Rev. Astron. Astrophys. {\bf 17}, 73.

\item SAO 1966 {\it Smithsonian Astrophysical Observatory Star Catalog}
Washington: Smithsonian Institution (SAO).


\item Schechter, P. 1976 Astrophys. J. {\bf 203}, 297.



\item Shull, J.M., Stocke, J.T. and Penton, S. 1996 Astron. J. {\bf 111}, 72.

\item Silva, D.R. and Cornell, M.E. 1992, Astrophys. J. Suppl. {\bf 81}, 865.

\item Smith, A.M. and Cornett, R.H. 1982 Astrophys. J. {\bf 261}, 1.


\item Storchi-Bergmann, T., Kinney, A. and Chalis, P. 1994, Astrophys. J. Suppl. 
{\bf 98}, 103.



\item Thompson, G.I., Nandy, K., Jamar, C. Monfils, A., Houziaux, L., Carnochan, D.J.
and Wilson, R. 1978 {\it Catalog of Stellar Ultraviolet Fluxes}, The Science Research
Council (TD-1).

\item Trinchieri, G., Fabbiano, G. and Canizares, C.R.C. 1986 Astrophys. J. {\bf 310}, 637.

\item Trinchieri, G., Kim, D.-W., Fabbiano, G. and Canizares, C.R.C. 1994 Astrophys. 
J. {\bf 428}, 555.


\item Turon, C. \etal 1993 Bull. Inf. CDS {\bf 43} (Version 2 of the Hipparcos
Input Catalog).


\item van den Bergh, S. 1977 Vistas in Astronomy {\bf 21}, 71.


\item Visvanathan, N. and Griersmith, D. 1979 Astrophys.J. {\bf 230}, 1.



\item Weymann, R., Rauch, M., Williams, R., Morris, S. and Heap, S. 1995 Astrophys. J. {\bf 438}, 650..


\item Wesselius, P.R., van Duinen, R.J., deJonse, A.R.W., Aalders, J.W.G., Luinse, W. and Wildeman, K.J.
 1982 Astron. Astrophys. Suppl. {\bf 49}, 427.


\end{description}

\section*{Figure captions}
\begin{itemize}
\item Figure 1: Combined image of the three FAUST fields toward the Virgo cluster. The UV sky
brightness was different at the time the images were obtained, thus the change in background 
appearance. M87 is near the center of the combined image.

\item Figure 2: Internal consistency check:  UV magnitudes for FAUST sources measured 
more than once. Straight lines connect measurements of the same source observed three times.
 
\item Figure 3a to 3d: Examples of spectra for typical Virgo cluster objects. Note that the
broad ``absorption'' near $\sim$4650\AA\, for V24 is an artefact originating from the
standard star observation of the specific night.

\item Figure 4: UV sky model  from Brosch (1991) calculated for
the FAUST spectral response (solid line), {\it vs.} the actual measured stellar densities
(squares).

\item Figure 5: Comparison of the HI flux density and the UV magnitude of galaxies. 
Early-type galaxies
are shown as triangles, intermediate types as squares, and late types as
diamonds. Open symbols indicate upper limits.

\item Figure 6: Comparison of the far infrared flux density (FIR) and the UV magnitude
of galaxies. Symbols are as in Fig. 5.

\item Figure 7: The change of the FIR/HI ratio as a function of (FAUST--B) for
galaxies.

\item Figure 8: The large-scale distribution of galaxies measured by FAUST and
with redshift available from the literature is shown in a diagram of right ascension 
{\it vs.} heliocentric velocity as a depth indicator. The cluster center is marked
by an asterisc at the velocity of M87.

\item Figure 9: UV luminosity function of Virgo galaxies (cluster and field galaxies
are included). The dashed line indicates the Gaussian fit to the observational data.

\end{itemize}


\newpage
\begin{deluxetable}{rrccclc}
\tablecaption{Detected FAUST sources in the Virgo images}
\tablehead{ \colhead{Id. no.} & \colhead{$\alpha$} &
\colhead{$\delta$} & \colhead{m$_{UV}$} &
\colhead{N$_{obs}$} & \colhead{Proposed id.} &
\colhead{FAUST no.}} 
\startdata
1 & *12 09 35.2 & 10 23 07 & 12.79$\pm$0.43 & 1 & No ID & - \nl
2 & *12 10 11.7 & 11 08 26 & 11.61$\pm$0.19 & 1 & N4178 & 3038  \nl
3 & *12 10 50.7 & 10 31 46 & 8.40$\pm$0.16 & 1 & SAO099997 & 3043  \nl
4 & 12 11 17.0 & 09 15 02 & 8.83$\pm$0.016 & 1 & SAO119277 & 3046 \nl
5 & *12 12 28.4 & 10 15 39 & 12.69$\pm$0.25 & 1 & SAO100006 & 3049 \nl
6 & 12 13 16.0 & 09 18 45 & 11.06$\pm$0.17 & 1 & SAO119294 & 3056 \nl
7 & 12 13 52.8 & 07 37 58 & 10.70$\pm$0.17 & 1 & SAO119300 & 3059 \nl
8 & 12 13 53.3 & 15 06 35 & 6.10$\pm$0.16 & 1 & SAO100012 & 3058 \nl
9 & *12 14 24.8 & 08 16 41 & 9.84$\pm$0.16 & 1 & S121426+081714 & 3061 \nl
10 & *12 14 56.3 & 06 53 18 & 11.65$\pm$0.23 & 1 & SAO119308 & 3065 \nl
11 & *12 15 01.6 & 15 49 03 & 10.42$\pm$0.18 & 1 & S121449+155138 & 3066 \nl
12 & *12 15 05.7 & 08 44 11 & 13.03$\pm$0.32 & 1 &  IC3111 & 3068 \nl
13 & *12 15 10.8 & 09 54 10 & 12.67$\pm$0.24 & 1 & S121508+095341 & 3069 \nl
14 & 12 15 17.0 & 10 23 51 & 7.62$\pm$0.16 & 2 & SAO100024 & 3072 \nl
15 & *12 15 25.6 & 07 27 27 & 12.83$\pm$0.35 & 1 & N4246 & 3074 \nl
16 & 12 15 40.0 & 10 08 46 & 9.86$\pm$0.77 & 2 & SAO100029 & 3078 \nl
17 & *12 16 12.3 & 14 40 53 & 10.08$\pm$0.17 & 2 & N4254 & 3079 \nl
18 & *12 16 17.1 & 10 23 11 & 13.40$\pm$0.26 & 1 & SAO100034 & 3080 \nl
19 & *12 16 28.5 & 13 12 47 & 11.94$\pm$0.23 & 2 & HD107116 & 3082 \nl
20 & *12 16 32.6 & 09 08 01 & 13.42$\pm$0.31 & 1 & IC0776 & 3083 \nl
21 &  12 16 52.0 & 07 16 10 & 9.43$\pm$0.16 & 1 & SAO119328 & 3085 \nl
22 & *12 17 15.5 & 08 57 13 & 12.94$\pm$0.26 & 1 & HD107227 & - \nl
23 & *12 17 35.5 & 07 57 52 & 12.90$\pm$0.31 & 1 & N4276 & 3089 \nl
24 & *12 17 51.2 & 10 14 50 & 8.92$\pm$0.16  & 2 & S121757+101353 & 3090/1 \nl
25 & *12 18 36.4 & 09 28 08 & 13.63$\pm$0.41 & 1 & S121836+092743 & 3093 \nl
26 & *12 18 48.5 & 11 48 15 & 11.57$\pm$0.17 & 2 & N4294 & 3094/6\nl
27 & *12 18 57.8 & 10 41 17 & 12.41$\pm$0.72 & 2 & S121860+104153 & 3098 \nl
28 & *12 19 01.1 & 11 47 58 & 11.58$\pm$0.17 & 2 & N4299 & 3099 \nl
29 & *12 19 01.8 & 14 53 27 & 12.21$\pm$0.31 & 2 & N4298 & 3097 \nl
30 & 12 19 30.0 & 17 01 20 & 6.81$\pm$0.16 & 1 & SAO100062 & 3101 \nl
31 & *12 20 23.6 & 16 04 55 & 10.07$\pm$0.17 & 2 & N4321 & 3104/5\nl
32 & *12 20 32.8 & 10 55 55 & 13.73$\pm$0.97 & 2 & N4325 & 3106 \nl
33 & *12 20 43.4 & 13 56 08 & 11.12$\pm$0.20 & 3 & SAO100071 & 3108 \nl
34 & *12 20 47.3 & 11 38 44 & 13.28$\pm$0.35 & 1 & N4330 & 3109 \nl
35 & *12 21 15.8 & 12 48 05 & 12.46$\pm$0.24 & 2 & IC3258 & 3111 \nl
36 & 12 21 18.0 & 08 37 17 & 11.38$\pm$0.16 & 1 & SAO119371 & 3113  \nl
37 & *12 21 28.7 & 06 27 45 & 12.71$\pm$0.26 & 1 & S122129+062746 & -\nl
38 & *12 21 34.1 & 12 31 10 & 12.77$\pm$0.38 & 2 & N4351 & 3114 \nl
39 & *12 21 36.4 & 06 52 47 & 12.60$\pm$0.25 & 1 & IC3268 & 3115 \nl
40 & *12 21 52.6 & 06 30 43 & 12.67$\pm$0.25 & 1 & S122154+063148 & 3116  \nl
41 & *12 21 58.0 & 07 35 46 & 13.98$\pm$0.61 & 1 & N4365 & 3117 \nl
42 & *12 22 11.8 & 08 48 15 & 14.10$\pm$0.60 & 1 & VCC 740 & 3118 \nl
43 & *12 22 27.0 & 11 55 45 & 13.84$\pm$0.44 & 1 & N4371 & - \nl
44 & *12 22 39.9 & 13 12 19 & 13.56$\pm$0.63 & 1 & N4374 ? & 3119 \nl
45 & *12 22 50.9 & 10 19 07 & 12.81$\pm$0.38 & 2 & N4380 & 3122 \nl
46 & *12 22 54.8 & 16 45 08 & 11.75$\pm$0.21 & 1 & N4383 & 3125 \nl
47 & *12 23 11.5 & 07 29 20 & 12.93$\pm$0.24 & 1 & IC3322A & 3127 \nl
48 & *12 23 19.4 & 10 45 46 & 12.73$\pm$0.40 & 2 & N4390 & 3129/31\nl
49 & *12 23 20.7 & 12 59 07 & 12.35$\pm$0.27 & 2 & N4388 & 3128 \nl
50 & *12 23 31.2 & 15 57 42 & 12.42$\pm$0.45 & 2 & N4396 & 3132 \nl
51 & *12 23 49.3 & 13 15 47 & 13.16$\pm$0.79 & 3 & N4406 & 3133 \nl
52 & *12 24 05.3 & 12 57 21 & 12.07$\pm$0.31 & 3 & N4413 or HD108285  ? & 3134 \nl
53 & *12 24 09.6 & 09 07 19 & 11.32$\pm$0.23 & 1 & N4411A & 3136 \nl
54 & *12 24 12.7 & 09 10 28 & 11.85$\pm$0.17 & 1 & N4411B & 3137 \nl
55 & *12 24 19.3 & 08 12 43 & 12.41$\pm$0.20 & 1 & N4416 & 3139 \nl
56 & *12 24 35.9 & 08 52 31 & 13.66$\pm$0.25 & 1 & No ID & 3140 \nl
57 & *12 24 41.8 & 07 33 17 & 12.27$\pm$0.17 & 1 & VCC 975 ? & 3142 \nl
58 & *12 24 42.4 & 09 42 49 & 14.16$\pm$0.59 & 1 & N4424 & 3144 \nl
59 & *12 24 44.3 & 16 10 34 & 12.77$\pm$0.64 & 2 & IC3365 & 3143 \nl
60 & 12 24 49.0 & 07 25 30 & 12.14$\pm$0.18& 1 & SAO119404 & 3145 \nl
61 & 12 24 52.0 & 07 39 31 & 9.55$\pm$0.16 & 1 & SAO119405 & 3147 \nl
62 & *12 24 57.0 & 08 54 44 & 13.50$\pm$0.26 & 1 & No ID & 3151 \nl
63 & *12 24 57.5 & 06 32 02 & 12.02$\pm$0.21 & 1 & N4432 & 3148 \nl
64 & *12 25 02.8 & 11 35 35 & 11.11$\pm$0.20 & 3 & HD108452 & 3150 \nl
65 & *12 25 15.2 & 11 55 25 & 11.23$\pm$0.24 & 3 & No ID & 3154 \nl
66 & *12 25 17.3 & 13 20 26 & 12.38$\pm$0.39 & 3 & N4438 & 3152/3/5 \nl
67 & 12 25 27.0 & 09 37 44 & 12.21$\pm$0.49 & 2 & SAO119416 & 3156/7 \nl
68 & *12 25 41.2 & 15 52 29 & 13.01$\pm$0.72 & 2 & SAO100112 & 3158 \nl
69 & *12 25 51.4 & 09 01 34 & 12.69$\pm$0.23 & 1 & VCC 1091 ? & 3159 \nl
70 & *12 25 52.4 & 12 13 04 & 15.46:: & 3 & S122551+121247 & - \nl
71 & *12 26 12.7 & 09 33 03 & 13.77$\pm$0.46 & 1 & N4451 & 3162 \nl
72 & *12 26 16.6 & 07 26 10 & 9.12$\pm$0.16 & 1 & S122614+072530 & 3164 \nl
73 & *12 26 30.9 & 05 57 10 & 13.29$\pm$0.76 & 1 & SAO119423 & 3165 \nl
74 & *12 26 32.0 & 10 58 46 & 13.03$\pm$0.52 & 2 & S122644+105941 & 3166 \nl
75 & *12 26 33.8 & 06 15 13 & 11.89$\pm$0.21 & 1 & SAO119424 & 3167 \nl
76 & *12 26 45.3 & 11 58 58 & 13.85$\pm$0.98 & 2 & PG 1226+119 & 3168 \nl
77 & 12 26 49.0 & 17 35 56 & 7.35$\pm$0.16 & 1 & SAO100122 & 3169 \nl
78 & *12 26 59.9 & 07 03 11 & 13.11$\pm$0.31 & 1 & IC3414 & 3174 \nl
79 & *12 27 01.3 & 06 31 44 & 12.64$\pm$0.26 & 1 & SAO119429 & 3173 \nl
80 & 12 27 00.0 & 11 46 06 & 11.97$\pm$0.33 & 3 & SAO100126 & 3172 \nl
81 & *12 27 08.5 & 08 07 08 & 12.13$\pm$0.19 & 1 & N4470 & 3177 \nl
82 & *12 27 12.4 & 08 37 48 & 11.12$\pm$0.24 & 2 & S122710+083901 & 3179/80 \nl
83 & *12 27 18.0 & 08 17 31 & 12.17$\pm$0.19 & 1 & N4472 & 3181 \nl
84 & *12 27 24.4 & 13 06 02 & 13.05$\pm$0.67 & 3 & S122720+130503 & 3182 \nl
85 & *12 28 04.3 & 10 01 23 & 12.80$\pm$0.36 & 2 & SAO 119436 & 3188 \nl
86 & *12 28 09.3 & 08 49 02 & 12.50$\pm$0.20 & 1 & SAO119438 & 3189 \nl
87 & *12 28 11.6 & 09 02 23 & 12.34$\pm$0.19 & 1 & S122810+090112 & 3190 \nl
88 & *12 28 18.4 & 12 41 15 & 12.02$\pm$0.30 & 3 & N4486 & 3191 \nl
89 & *12 28 45.9 & 05 21 17 & 9.76$\pm$0.16 & 1 & SAO119454 & 3197 \nl
90 & *12 28 57.5 & 11 48 40 & 13.45$\pm$1.28 & 3 & IC3446 & 3200 \nl
91 &*12 29 06.1 & 17 08 33 & 12.09$\pm$0.26 & 1 & N4489 & 3207 \nl
92 & *12 29 07.1 & 08 03 37 & 13.90$\pm$0.38 & 1 & No ID & 3204 \nl
93 & 12 29 07.0 & 12 24 13 & 11.02$\pm$0.19 & 3 & SAO100140 & 3206 \nl
94 & *12 29 20.5 & 15 24 18 & 13.13$\pm$0.65 & 2 & IC0797 & 3214 \nl
95 & *12 29 27.4 & 14 42 34 & 11.10$\pm$0.18 & 2 & N4501 & 3211 \nl
96 & *12 29 35.8 & 12 34 10 & 13.07$\pm$0.44 & 2 & S122936+123224 & 3216  \nl
97 & *12 29 38.7 & 08 02 36 & 12.92$\pm$0.25 & 1 & S122935+080140 & - \nl
98 & *12 29 47.4 & 06 00 28 & 12.35$\pm$0.28 & 1 & SAO119459 & 3219 \nl
99 & *12 29 55.8 & 09 41 09 & 13.10$\pm$0.80 & 2 & No ID & 3221 \nl
100 & *12 30 12.9 & 14 20 47 & 12.07$\pm$0.38 & 3 & IC3476 & 3222 \nl
101 & *12 30 16.9 & 09 50 33 & 13.89$\pm$0.39 & 1 & No ID & 3224 \nl
102 & *12 30 29.4 & 13 13 11 & 13.13$\pm$0.73 & 3 & S123030+131240 & 3228 \nl
103 & *12 30 30.4 & 08 12 41 & 12.49$\pm$0.21 & 1 & S123028+081156 & 3226 \nl
104 & *12 30 34.3 & 05 18 07 & 9.43$\pm$0.16 & 1 & No ID & - \nl
105 & *12 30 51.7 & 06 41 56 & 8.74$\pm$0.16 & 1 & PG 1230+067 & 3231  \nl
106 & *12 30 54.0 & 09 49 10 & 13.92$\pm$1.20 & 2 & Q 1230+0947 & 3233 \nl
107 & *12 30 59.5 & 08 55 05 & 11.13$\pm$0.23 & 2 & N4519/4519a & 3235/6 \nl
108 & *12 31 05.8 & 09 11 50 & 13.81$\pm$0.52 & 1 & SAO119467 & 3237 \nl
109 & *12 31 07.9 & 09 26 30 & 12.26$\pm$0.45 & 2 & N4522 & 3238 \nl
110 & *12 31 13.7 & 15 25 36 & 11.87$\pm$0.30 & 2 & N4523 & 3240 \nl
111 & *12 31 21.8 & 08 18 11 & 14.17$\pm$0.63 & 1 & VCC 1525 & - \nl
112 & * 12 31 37.0 & 17 13 01 & 11.53$\pm$0.20 & 1 & No ID & 3248 \nl
113 & *12 31 46.9 & 06 44 51 & 10.95$\pm$0.16 & 1 & N4532 & 3249 \nl
114 & *12 31 49.4 & 08 19 13 & 10.18$\pm$0.16 & 1 & N4535/4535a & - \nl
115 & 12 31 48.0 & 11 11 46 & 10.91$\pm$0.18 & 3 & SAO100149 & 3250 \nl
116 & *12 32 00.4 & 15 45 23 & 12.81$\pm$0.52 & 2 &  No ID  & - \nl
117 & 12 32 20.0 & 11 06 43 & 12.74$\pm$0.18 & 3 & SAO100156 & 3258 \nl
118 & *12 32 26.7 & 15 50 25 & 13.16$\pm$0.36 & 1 & IC3528 & 3256 \nl
119 & *12 32 34.0 & 13 36 32 & 12.65$\pm$0.48 & 3 & No ID & 3260 \nl
120 & 12 32 37.0 & 18 39 07 & 8.92$\pm$0.16& 1 & SAO100159 & 3260 \nl
121 & *12 32 41.3 & 10 48 22 & 13.32$\pm$0.52 & 2 & VCC 1600 ? & 3263 \nl
122 & *12 32 53.3 & 09 44 57 & 10.49$\pm$0.16 & 2 & S123252+094327 & 3268 \nl
123 & *12 32 55.9 & 14 46 40 & 11.90$\pm$0.27 & 2 & N4548 & 3267 \nl
124 & *12 33 15.4 & 12 51 41 & 13.10$\pm$0.68 & 3 & N4552 & - \nl
125 & *12 33 59.5 & 07 54 46 & 13.41$\pm$0.37 & 1 & No ID & 3278 \nl
126 & *12 34 03.7 & 11 33 19 & 11.57$\pm$0.26 & 3 & N4567 or N4568 & 3280  \nl
127 & *12 34 05.0 & 06 53 07 & 12.63$\pm$0.25 & 1 & IC3576 & 3279 \nl
128 & 12 34 13.0 & 15 32 21 & 12.11$\pm$0.33 & 2 & SAO100173 & 3287 \nl
129 & *12 34 16.4 & 13 28 44 & 11.38$\pm$0.21 & 3 & N4569 & 3281 \nl
130 & 12 34 17.0 & 08 12 30 & 12.37$\pm$0.19 & 1 & SAO119494 & 3284 \nl
131 & 12 34 31.0 & 14 31 38 & 11.86$\pm$0.25 & 2 & N4571 or SA0100177 & 3286 \nl
132 & *12 34 33.8 & 07 11 44 & 12.99$\pm$0.31 & 1 & IC3591 & 3288 \nl
133 & 12 34 35.0 & 09 04 19 & 9.28$\pm$0.16 & 2 & SAO119497 & 3290 \nl
134 & *12 35 09.4 & 10 04 44 & 12.25$\pm$0.30 & 2 & S123507+100344 & 3299 \nl
135 & *12 35 10.8 & 15 50 16 & 10.73$\pm$0.18 & 2 & S123515+155104 & 3297 \nl
136 & *12 35 11.7 & 08 50 27 & 13.10$\pm$0.27 & 1 & VCC 1725 & 3298 \nl
137 & *12 35 14.2 & 07 22 44 & 13.62$\pm$ 0.51 & 1 & U7795 & 3300 \nl
138 & *12 35 14.9 & 12 07 11 & 12.03$\pm$0.32 & 3 & N4579 & 3302 \nl
139 & *12 36 00.0 & 10 32 33 & 14.33:: & 2 & S123558+103139 & - \nl
140 & *12 36 24.6 & 16 41 21 & 11.47$\pm$0.26 & 2 & SAO100194 & 3311 \nl
141 & *12 36 29.1 & 14 03 36 & 11.16$\pm$0.19 & 2 & SAO100192 & 3319 \nl
142 & 12 36 32.0 & 17 42 42 & 11.62$\pm$0.23 & 1 & SAO100193 & 3323 \nl
143 & 12 36 39.0 & 14 36 06 & 11.72$\pm$0.23 & 2 & SAO100195 & 3322 \nl
144 & *12 36 53.3 & 08 14 14 & 13.17$\pm$0.32 & 1 & IC3617 & 3331 \nl
145 & *12 37 09.5 & 13 14 25 & 10.42$\pm$0.17 & 3 & PG 1237+132 & 3336 \nl
146 & *12 37 19.7 & 15 33 59 & 12.57$\pm$0.42 & 2 & N4595 & 3339 \nl
147 & *12 37 34.8 & 10 26 31 & 12.50$\pm$1.04 & 2 & N4596 & 3342 \nl
148 & *12 37 35.1 & 11 49 24 & 11.80$\pm$0.28 & 3 & PG 1237+118 & 3347 \nl
149 & *12 37 41.6 & 15 05 17 & 12.72$\pm$0.47 & 2 & S123734+150560 & 3343 \nl
151 & *12 37 49.7 & 09 24 00 & 13.68$\pm$0.47 & 1 & SAO119524 & 3352 \nl
152 & *12 37 55.5 & 10 15 57 & 11.76$\pm$0.26 & 2 & S123754+101551 & 3355 \nl
150 & 12 37 59.0 & 16 33 12 & 12.50$\pm$0.19 & 2 & SAO100201 & 3357 \nl
153 & 12 38 22.0 & 09 06 13 & 11.50$\pm$0.17 & 1 & SAO119530 & 3360 \nl
154 & *12 38 34.8 & 07 15 16 & 10.49$\pm$0.16 & 1 & No ID & 3365 \nl
155 & *12 38 39.5 & 08 28 07 & 13.04$\pm$0.32 & 1 & KARA 68181 ? & 3367 \nl
156 & *12 38 52.3 & 09 58 23 & 13.08$\pm$0.36 & 1 & S123853+095836 & - \nl
157 & *12 38 52.3 & 10 53 13 & 12.41$\pm$0.21 & 1 & S123847+105318 & 3371 \nl
158 & 12 39 03.0 & 10 42 02 & 7.63$\pm$0.16 & 2 & SAO100207 & 3375 \nl
159 & 12 39 17.0 & 10 09 36 & 9.76$\pm$0.16 & 2 & SAO100210 & 3380 \nl
160 & 12 39 21.0 & 10 30 39 & 5.54$\pm$0.16 & 2 & SAO100211 & 3382 \nl
161 & 12 39 25.0 & 07 04 51 & 6.17$\pm$0.16 & 1 & SAO119538 & 3381 \nl
162 & *12 39 27.0 & 17 47 19 & 6.95$\pm$0.16 & 1 & PG 1239+178 & 3385 \nl
163 & *12 39 31.0 & 08 55 18 & 14.10$\pm$0.84 & 1 & S123934+085455 & - \nl
164 & *12 39 37.5 & 16 37 19 & 12.10$\pm$0.36 & 2 & TD1 26442 ? & 3389 \nl
165 & 12 39 39.0 & 11 09 26 & 11.08$\pm$0.19 & 2 & SAO100213 & 3388 \nl
166 & *12 39 42.3 & 08 18 15 & 13.68$\pm$0.55 & 1 & SAO119543 & 3387 \nl
167 & *12 40 02.2 & 12 00 58 & 11.14$\pm$0.23 & 3 & PG 1240+120 & 3394 \nl
168 & *12 40 13.6 & 14 35 59 & 12.54$\pm$0.44 & 2 & N4634 & 3393 \nl
169 & *12 40 23.6 & 13 53 25 & 13.25$\pm$0.72 & 2 &  S124024+135325& 3399 \nl
170 & *12 40 24.7 & 13 32 07 & 11.71$\pm$ 0.24 & 2 & N4639 & 3400 \nl
171 & *12 41 07.6 & 11 50 42 & 11.40$\pm$0.20 & 2 & N4649 or N4647 & 3409 \nl
172 & *12 41 16.6 & 16 40 09 & 10.97$\pm$0.18 & 1 & N4651 & 3414 \nl
173 & 12 41 20.8 & 15 23 09 & 11.51$\pm$0.25 & 2 & SAO100228 & 3415 \nl
174 & *12 41 28.9 & 13 24 24 & 10.86$\pm$0.18 & 2 & N4654 & 3417 \nl
175 & 12 41 30.0 & 11 11 42 & 11.05$\pm$0.19 & 2 & SAO100229 & 3419 \nl
176 & *12 41 57.0 & 10 37 21 & 11.93$\pm$0.30 & 2 & S124155+103724 & 3426  \nl
177 & *12 43 01.6 & 14 39 18 & 5.57$\pm$0.16 & 2 & SAO100235+100236 & 3441 \nl
178 & *12 43 01.9 & 07 56 12 & 8.65$\pm$0.16 & 1 & SAO119574=FM Vir & 3442 \nl
179 & *12 43 070 & 13 37 43 & 12.72$\pm$0.55 & 2 & IC3742 & - \nl
180 & *12 43 10.0 & 10 35 35 & 10.74$\pm$0.19 & 1 & No ID & - \nl
181 & *12 43 50.4 & 09 32 03 & 10.66$\pm$0.17 & 1 & SAO119582 & 3454 \nl
182 & 12 44 08.8 & 16 51 01 & 10.45$\pm$0.17 & 1 & SAO100249/252/257 & 3456 \nl
183 & *12 44 31.8 & 11 19 14 & 10.05$\pm$0.17 & 2 & PG 1244+113 & 3460 \nl
184 & 12 44 42.6 & 12 13 52 & 6.74$\pm$0.16 & 1 & SAO100260 & 3461 \nl
185 & *12 44 42.3 & 13 53 47 & 11.54$\pm$0.36 & 2 & SAO100258 & 3463 \nl
186 & *12 45 11.8 & 13 35 55 & 11.98$\pm$0.22 & 1 & S124507+133509 & 3469 \nl
\tablebreak
187 & *12 45 21.6 & 14 02 13 & 11.59$\pm$0.31 & 2 & N4689 & 3473 \nl
188 & *12 45 41.0 & 09 21 21 & 11.30$\pm$0.21 & 1 & SAO119602 & 3480 \nl
189 & 12 45 44.1 & 13 49 33 & 6.23$\pm$0.16 & 2 & SAO100269 & 3483 \nl
190 & *12 46 10.8 & 12 26 28 & 9.10$\pm$0.16 & 1 & S124615+122650 & 3491  \nl
191 & 12 46 23.8 & 14 23 43 & 7.20$\pm$1.54 & 2 & SAO100283 & 3497 \nl
\enddata
\tablecomments{1. The adopted UV magnitude is the average of the different FAUST 
measurements. The error is the root-mean-square of the the individual errors.
2. The : indicates a very large measurement error.
3.  N$_{obs}$ is the number of times a source was observed, \ie the number of FAUST
images in which it appears.
4.  The question mark ? at  N$ 44 $, N$ 57 $, and N$ 69 $ indicates that the
 proposed identification is a bit off the  FAUST source.
5.  N$ 52 $: The UV source may be a blend of the emission from the galaxy
and the  HD108285 (A2) star.
6.  N$ 155 $: The three stars near the  FAUST source have been observed spectroscopically
and identified with late G  or K star. The proposed identification is the faint KARA 68181 galaxy  marked on the overlays of the PSS.
Note that there is a dE galaxy (VCC 1866) a bit far from the FAUST source. 
7.  N$ 164 $: This source is near the TD-1 26442 from the Extended TD-1 catalog (preprint).
On the PSS there is a faint V$ > $ 15 blue looking candidate 123945+163743 not observed. }

\end{deluxetable}

\pagebreak

\begin{deluxetable}{rcrrrrrr}
\tablecaption{Stars towards the Virgo cluster}
\small
\tablehead{
\colhead{Source no.} & \colhead{ID} & 
\colhead{Sp. Type} & \colhead{V} & 
\colhead{$B-V$} & \colhead{$U-B$} & 
\colhead{FAUST} & \colhead{Other IDs}}
\startdata
3 & SAO099997    & A2m & 5.8 & 0.26 & .09 & 8.40$\pm$0.16 & HIC 59608 \nl
4 & SAO119277   & A2   & 7.6 & 0.16 &      & 8.83$\pm$0.16 & HIC 59636 \nl
5 & SAO100006   & F5   & 8.7 & 0.48 &      & 12.69$\pm$0.25 & HIC 59737 \nl
6 & SAO119294   & A2  & 8.85 & 0.28 &      & 11.06$\pm$0.17 & HIC 59804 \nl
7 & SAO119300   & A3  & 8.40 & 0.32 &      & 10.70$\pm$0.17 & HIC 59852 \nl
8 & SAO100012   & A3  & 5.10 & 0.06 &      & 6.10$\pm$0.17 & HIC 59819 \nl
9 & S121426+081714 &B(e) & 11.9  &   &      & 9.84$\pm$0.16 &  \nl
10 & SAO119308  & F2  & 7.9  & 0.40 &      & 11.65$\pm$0.23 & HIC 59916 \nl
11 & S121449+155138 & B(l) & 11.7 &    &     &10.42$\pm$0.18 & \nl
13 & S121508+095341 & A(l) & 14.4 &    &     & 12.67$\pm$0.24 &  \nl
14 & SAO100024 & B9  & 9.81  & --0.12 &      & 7.62$\pm$0.16 & HIC 59955 \nl
16 & SAO100029 & A3  & 8.89  & 0.17 &      & 9.86$\pm$0.77 & HIC 59992 \nl
18 & SAO100034 & F0  & 9.3   &   &      & 13.40$\pm$0.26 & HD107089 \nl
19 & HD107116  & A3  & 10.38 & 0.23 &      & 11.94$\pm$0.23 & HIC 60065 \nl
21 & SAO119328 & A0  & 8.60 & 0.09 &      & 9.43$\pm$0.16 & HIC 60095 \nl
22 & HD107227 & A3 & 10.8 &    &      &12.94$\pm$0.26 & \nl
24 & S121757+101353 & G(e) & 13.2 &    &   & 8.92$\pm$0.16 & \nl
25 & S121836+092743 & A(e) & 12.6 &   &    & 13.63$\pm$0.41 & \nl
27 & S121860+104153 & B(l) & 12.9 &     &   & 12.41$\pm$0.72 & \nl
30 & SAO100062 & A2 & 6.66 & 0.05 & .05 & 6.81$\pm$0.16 & HIC 60313 \nl
33 & SAO100071 & F5  & 8.55 & 0.33 &       & 11.12$\pm$0.20 & HIC 60419 \nl
36 & SAO119371 & F0 & 8.83 & 0.32 &       & 11.38$\pm$0.16 & HIC 60473 \nl
37 & S122129+062746 & A(e) & 12.5 &    &   & 12.71$\pm$0.26 & \nl
40 & S122154+063148 & B(l) & 12.2 &    &   & 12.67$\pm$0.25 & \nl
60 & SAO119404 & F2 & 8.43 & 0.44 &      & 12.14$\pm$0.18 & HIC 60766 \nl
61 & SAO119405 & A0 & 8.35 & 0.43 &      & 9.55$\pm$0.16 & HIC 60772 \nl
64 & HD108452  & A0 & 10.0 &     &      & 11.11$\pm$0.20 & \nl
67 & SAO119416 & F5  & 9.4 &      &      & 12.21$\pm$0.49 & HD108535 \nl
68 & SAO100112 & F8 & 8.6 & 0.42 &      & 13.01$\pm$0.72 & HIC 60836 \nl
70 & S122551+121247  & A(l) & $>$ 15 &     &      & 15.46:: & \nl
72 & S122614+072530 & A(l) & 10.8 &     &   & 9.12$\pm$0.16 & \nl
73 & SAO119423 & A2   & 8.7 &      &      &13.29$\pm$0.74 &  HD108663 \nl
74 & S122644+105941 & F(l) & 12.6 &  &   & 13.03$\pm$0.52 & \nl
75 & SAO119424 & F2  &  8.1 & 0.44 &   & 11.89$\pm$0.21 & HIC 60915 \nl
76 & PG 1226+119 & HBB & 15.37 (B) &  &   & 13.85$\pm$0.98 &        \nl
77 & SAO100122 & A0 & 7.7 & 0.11 & --.01 & 7.35$\pm$0.16 & HIC 60933 \nl
79 & SAO119429 & G0  & 8.7 &      &     & 12.64$\pm$0.26 & HD108738 \nl
80 & SAO100126 & F0   & 8.53 & 0.40 &     & 11.97$\pm$0.33 & HIC 60945 \nl
82 & S122710+083901 & B(l) & 11.5 &     &   & 11.12$\pm$0.24 &  \nl
84 & S122720+130503 & A(l) & 10.5 &     &  & 13.05$\pm$0.67 &  \nl
85 & SAO119436 & F2  & 7.5 &  0.48 &      &12.80$\pm$0.36 & HIC 61035 \nl
86 & SAO119438 & A5 & 8.8 &      &      &12.50$\pm$0.20 & HD108876 \nl
87 & S122810+090112 & A(l) & 11.2 &   &      & 12.34$\pm$0.19 & \nl
89 & SAO119454   & F2 & 7.8 & 0.36 &      & 9.76$\pm$0.16 & HIC 61104 \nl
93 & SAO100140   & F0  & 8.1 & 0.34 &      & 11.02$\pm$0.19 & HIC 61135 \nl
96 & S122936+123224 & A(e) & 10.9 &     &      & 13.07$\pm$0.43 & \nl
97 & S122935+080140 & A(e) & 11. &      &      & 12.92$\pm$0.25 & \nl
98 & SAO119459  & F0   & 8.9 &   &      & 12.35$\pm$0.28 & HD109129 \nl
102 & S123030+131240 & B(l) & 13.2 &    &      & 13.13$\pm$0.73 & \nl
103 & S123028+081156 & B(l) & 10.5 &    &      & 12.49$\pm$0.21 & \nl
105 & PG 1230+067 & sdO-B & 13.4 (B) &    &      & 8.74$\pm$0.16 &  \nl
108 & SAO119467 & F8 & 8.9 & 0.56 &      & 13.81$\pm$0.52 & HIC 61297 \nl
115 & SAO100149 & A5 & 7.6 & 0.40 &      & 10.91$\pm$0.18 & HIC 61353 \nl
117 & SAO100156  & F0 & 7.9 & 0.30 &      & 12.74$\pm$0.18 & HIC 61397 \nl
120 & SAO100159  & A7m & 6.6 & 0.27 &    &  8.92$\pm$0.16 & HIC 61415 \nl
122 & S123252+094327 & B(l) & 11.4 &    &      & 10.49$\pm$0.16 &  \nl
128 & SAO100173  &      & 8.7 &     &      & 12.11$\pm$0.33 & \nl
130 & SAO119494  & F8 & 8.9 & 0.56 &      & 12.37$\pm$0.19 & HIC 61561 \nl
133 & SAO119497  & A2 & 6.6 & 0.25 &      & 9.28$\pm$0.16 &  HIC 61579 \nl
134 & S123507+100344 & B(l) & 10.6 &    &      & 12.25$\pm$0.30 & \nl
135 & S123515+155104 & B(l) & 10.0 &    &      & 10.73$\pm$0.18 & \nl
139 & S123558+103139 & A(e) & 11.2 &    &      & 14.33:: & \nl
140 & SAO100194 & F3 & 7.6 & 0.45 &      & 11.47$\pm$0.26 & HIC 61727 \nl
141 & SAO100192 & F8 & 8.9 &       &      & 11.16$\pm$0.19 & HD109997  \nl
142 & SAO100193 & F0 & 9.45 & 0.31 &      & 11.62$\pm$0.23 & HIC 61718 \nl
143 & SAO100195 & A0m & 7.3 &  0.99 &      & 11.72$\pm$0.23 & HIC 61728 \nl
145 & PG 1237+132 & sdB & 13.96 (B) &   &  & 10.42$\pm$0.17 & TON 97 \nl
148 & PG 1237+118 &sdB-O & 15.7 & --.13 & --1.21 & 11.80$\pm$0.28 &TON 98 \nl
149 & S123734+150560 & A(e) & 13.5 &    &      & 12.72$\pm$0.47 & \nl
150 & SAO100201  & F5 & 7.9 & 0.39 &      & 12.50$\pm$0.19 & HIC 61844 \nl
151 & SAO119524  & F8 & 8.9 &       &      & 13.68$\pm$0.47 & HD110197 \nl
152 & S123754+101551  & A(e) & $>$ 15 &   &      & 11.76$\pm$0.26 & \nl
153 & SAO119530 & F2 & 7.2 & 0.43 &      & 11.50$\pm$0.17 & HIC 61889\nl
156 & S123853+095836 & A(e) & 12.7 &    &      & 13.08$\pm$0.36 & \nl
157 & S123847+105318 & A(l) & 12.2 &    &      & 12.41$\pm$0.21 & \nl
158 & SAO100207  & A7 & 6.2 & 0.19 &      &  7.63$\pm$0.16 & HIC 61937 \nl
159 & SAO100210 & F0 & 7.0 & 0.32  &      &  9.76$\pm$0.16 & HIC 61950 \nl
160 & SAO100211 & A0 & 4.9 & 0.08 &      &  5.54$\pm$0.16 & HIC 61960  \nl
161 & SAO119538  & A2 & 5.6 & 0.00 & -0.01 &  6.17$\pm$0.16 & HIC 61968 \nl
162 & PG 1239+178 & sdO & 10.95 & --.32 & --1.22 & 6.95$\pm$0.16 & BD+18$^{\circ}$2647 \nl
163 & S123934+085455 & A(l) & 12.2 &    &    & 14.10$\pm$0.84 & \nl
164 & TD1 26442   &   ?   &      &    &      & 12.10$\pm$0.36 & \nl
165 & SAO100213  & A0 & 8.8 &       &      & 11.08$\pm$0.19 & HD110466 \nl
166 & SAO119543   & F8 & 7.9 & 0.50 &      & 13.68$\pm$0.55 & HIC 61990 \nl
167 & PG 1240+120 & sdO-B & 15.3 (B) &  &  & 11.14$\pm$0.23 & Ton 109  \nl
169 & S124024+135325 & B &  $>$ 15 &    &    & 13.25$\pm$0.72 & \nl
173 & SAO100228  & F2 & 8.8 & 0.27 &      & 11.51$\pm$0.25 & HIC 62116 \nl
175 & SAO100229   & F2 & 8.6 & 0.32 &      & 11.05$\pm$0.19 & HIC 62133 \nl
176 & S124155+103724 & B(l) & 12.0 &    &      & 11.93$\pm$0.30 & \nl
177 & SAO100235+100236& A0 &  &        &      &  5.57$\pm$0.16 & \nl
178 & SAO119574   & A8m & 5.2 & 0.32 &      &  8.65$\pm$0.16 & FM Vir \nl
181 & SAO119582   & F0 & 7.7 &      &      & 10.66$\pm$0.17 & HD111043 \nl
182 & SAO100249/252/257 & F-K & 5 &    &  & 10.45$\pm$0.17 & HD111067 \nl
183 & PG 1244+113 & sd B & 13.28  & --.23 & --1.19 & 10.05$\pm$0.17 & \nl
184 & SAO100260  & A3 & 6.1 & 0.12 &       &  6.74$\pm$0.16 & HIC 62394 \nl
185 & SAO100258  & F5 & 8.2 &     &       & 11.54$\pm$0.36 & HD111155 \nl
186 & S124507+133509 & B(e) & 12.1 &   &      & 11.98$\pm$0.22 & \nl
188 & SAO119602 & F5 & 7.9 & 0.42 &       & 11.30$\pm$0.21 & HIC 62489 \nl
189 & SAO100269  & A1 & 6.5 & 0.02 &       &  6.23$\pm$0.16 & HIC 62478 \nl
190 & S124615+122650 & B(l) & 10.4 &  &       &  9.10$\pm$0.16 & \nl
191 & SAO100283  & A1 & 5.7 & 0.02 &   &  7.20$\pm$1.54 & HIC 62541 \nl
\enddata
\end{deluxetable}

\pagebreak

\begin{deluxetable}{rcccccccc}
\tablecaption{Distribution of UV sources}
\tablehead{\colhead{$m_{UV}$} & \colhead{$B0-B9$} & 
\colhead{A0$-$A9} & \colhead{F0$-$F9} & 
\colhead{G0$-$G9} & \colhead{HBB, sdO/B} & 
\colhead{AGN/Galaxies} & \colhead{No ID} & 
\colhead{No Sp}} 
\startdata
5--6 & 0 & 2 & 0 & 0 & 0 & 0 & 0 & 0 \nl
6--7 & 0 & 5 & 0 & 0 & 1 & 0 & 0 & 0 \nl
7--8 & 1 & 3 & 0 & 0 & 0 & 0 & 0 & 0 \nl
8--9 & 0 & 4 & 0 & 1 & 1 & 0 & 0 & 0 \nl
9--10 & 2 & 5 & 2 & 0 & 0 & 0 & 1 & 0 \nl
10--11 & 3 & 2 & 2 & 0 & 2 & 6 & 2 & 0 \nl
11--12 & 3 & 6 & 14 & 0 & 2 & 16 & 2 & 0 \nl
12--13 & 4 & 8 & 8 & 1 & 0 & 31 & 3 & 2 \nl
13--14 & 2 & 5 & 6 & 0 & 1 & 20 & 6 & 0 \nl
14--15 & 0 & 2 & 0 & 0 & 0 & 3 & 0 & 0 \nl
$>$15 & 0 & 1 & 0 & 0 & 0 & 0 & 0 & 0 \nl
Total & 15 & 43 & 32 & 2 & 7 & 76 & 14 & 2 \nl
\enddata

\tablecomments{Sources marked "No Sp." are tentatively identified with objects
on the PSS, which have not been observed spectroscopically.}

\end{deluxetable}

\pagebreak

\begin{deluxetable}{rrlrrrrr}
\tablecaption{Galaxies towards the Virgo cluster-UV data}
\small
\tablehead{\colhead{Source no.} & 
\colhead{Name} & \colhead{Type} & \colhead{FAUST}& \colhead{4xE(B-V)}  
& \colhead{GUV} & \colhead{SCAP} & \colhead{S+C}}
\startdata
2 & N4178 & SBc(s)II &-.00& 11.61$\pm$0.19 & &          & 12.34 \nl
12 & IC3111 & Sb(s)II&-.05  & 13.03$\pm$0.32 & & &  \nl
15 & N4246 & Sc(s)I-II &-.06& 12.83$\pm$0.35 & & &  \nl
17 & N4254 & Sc(s)I.3 &.14& 10.08$\pm$0.17 &  & 10.1 & 10.74   \nl
20 & IC0776 & SBcd(s)III &-.03& 13.42$\pm$0.31 &  &  &   \nl
23 & N4276 & Sc(s)II &-.02& 12.90$\pm$0.31 &  &  & 15.82  \nl
26 & N4294 & SBc(s)II-III &.02& 11.57$\pm$0.17 &  &  &  13.07\nl
28 & N4299 & Scd(s)III&.02& 11.58$\pm$0.17 &  &  & 12.92  \nl
29 & N4298 & Sc(s)III &.10& 12.21$\pm$0.31 &  &  & 12.94   \nl
31 & N4321 & Sc(s)I &.04& 10.07$\pm$0.17 &  & 10.1 & 10.54   \nl
32 & N4325 & E4 & -.01&13.73$\pm$0.97 &  &  & $>$16.30 \nl
34 & N4330 & Sd(on edge)&.03 & 13.28$\pm$0.35 &  &  & 14.39 \nl
35 & IC3258 & ScIII-IV &.07& 12.46$\pm$0.24 & 14.69$\pm$0.75 &  & 13.37  \nl
38 & N4351 & Sc(s)II.3&.01& 12.77$\pm$0.38 & 13.16$\pm$0.20 & & 13.97  \nl
39 & IC3268 & Sc(s)III-IV or Sm & -.02&12.60$\pm$0.25 &  &  &   \nl
41 & N4365 & E3&-.03 & 13.98$\pm$0.61 &  &  & 13.64 \nl
42 & VCC 740 & SBmIII&-.03&  14.10$\pm$0.60 & & &  \nl
43 & N4371 &  $\rm SB0_{2}$(r)(3)&.02 & 13.84$\pm$0.44 & $>$14.58 &  & 15.39 \nl
44 & N4374 ? & E1&.13 & 13.56$\pm$0.63 & 14.38$\pm$0.6 &  & 12.59  \nl
45 & N4380 & Sab(s)&-.01 & 12.81$\pm$0.38 & 13.17$\pm$0.22 &  & 13.97 \nl
46 & N4383 & Amorphous &.02& 11.75$\pm$0.21 &  & 12.1 & 12.59  \nl
47 & IC3322A & Sc(on edge)&-.02 & 12.93$\pm$0.24 &  &  & 14.79   \nl
48 & N4390 & Sbc(s)II&.00& 12.73$\pm$0.40 & 13.70$\pm$0.35 & & 13.62   \nl
49 & N4388 & Sab &.11& 12.35$\pm$0.27 & 12.83$\pm$0.18 & & 13.42  \nl
50 & N4396 & Sc(s)II &.04& 12.42$\pm$0.45 &  & 12.6 & 13.04  \nl
51 & N4406 & $\rm S0_{1}$(3),E3 &.11& 13.16$\pm$0.79 & $>$14.82 & & 13.22\nl
52 & N4413 ? & SBbc(rs)II-III&.11& 12.07$\pm$0.31 &13.11$\pm$0.20 & & 14.32 \nl
53 & N4411A & SBc(s)II &-.03& 11.32$\pm$0.23 & & & 15.39 \nl
54 & N4411B & Sc(s)II &-.02& 11.85$\pm$0.17 & 9.90$\pm$0.13 & & 13.49  \nl
55 & N4416 & SBc(s)II.2 &-.02&12.41$\pm$0.20 & 14.08$\pm$0.39 & & $>$16.30 \nl
57 & VCC 975 ? & Scd(s)II&-.02 & 12.27$\pm$0.17 & & &  \nl
58 & N4424 & Sa pec &-.03& 14.16$\pm$0.59 & 13.61$\pm$0.31 & & 14.14  \nl
59 & IC3365 & Scd(s)III&.03 & 12.77$\pm$0.64 & & & \nl
63 & N4432 & Sc(s)I-II&.12 & 12.02$\pm$0.21 & & &\nl
66 & N4438 & Sb(tides)&.10 & 12.38$\pm$0.39 & 13.83$\pm$0.31 & & 13.49 \nl
69 & VCC 1091 ? & Sbc(s)I.8 &-.03& 12.69$\pm$0.23 & & & \nl
71 & N4451 &  Sc(s)III&-.03 & 13.77$\pm$0.46 & $>$14.82 & & 14.64 \nl
78 & IC3414 & Sc(s)II&-.03 & 13.11$\pm$0.31 & & & \nl
81 & N4470 & ScIII pec&-.02 & 12.13$\pm$0.19 & 12.60$\pm$0.13 & & 13.84\nl
83 & N4472 & E2/$\rm S0_{1}$(2)&-.02 & 12.17$\pm$0.19 & 13.01$\pm$0.20 & & 12.47\nl
88 & N4486 & E0 &.09&12.02$\pm$0.30 & 12.24$\pm$0.12 & & 12.68 \nl
90 & IC3446 & SmIII/BCD &.03& 13.45$\pm$1.28 & & & \nl
91 & N4489 & $\rm S0_{1}$(1)&.03& 12.09$\pm$0.26 & & &\nl
94 & IC0797 & SBc(s)II-III&.09 & 13.13$\pm$0.65 & & &  \nl
95 & N4501 & Sbc(s)II &.09& 11.10$\pm$0.18 & 12.87$\pm$0.22 &  & 11.29 \nl
100 & IC3476 & Sc(s)II.2&.09 & 12.07$\pm$0.38 & 12.87$\pm$0.30 & & 13.64 \nl
106 & Q 1230+0947 &  &-.03& 13.92$\pm$1.29 & &  &  \nl
107 & N4519/4519a & SBc(rs)II.2, dSO?&-.02 & 11.13$\pm$0.16 & 11.59$\pm$0.07 & & 13.04 \nl
109 & N4522 & Sc/Sb &-.04& 12.26$\pm$0.45 & 13.52$\pm$0.24 & & 14.29\nl
110 & N4523 & SBd(s)III&.09 & 11.87$\pm$0.30 & & & 13.07 \nl
111 & VCC 1525 & Sbc(r) II&-.01 &  14.17$\pm$0.63 & & & \nl
113 & N4532 & SmIII &-.04& 10.95$\pm$0.16 & & & \nl
114 & N4535/4535a & S, SBc(s)I.3 &-.01& 10.18$\pm$0.16 & 10.90$\pm$0.04 & & 11.39 \nl
118 & IC3528 & Sbc &.04& 13.16$\pm$0.36 & & &  \nl
121 & VCC 1600 ? & dE & -.01 &13.32$\pm$0.52 & & & \nl
123 & N4548 & SBb(rs)I-II &.07& 11.90$\pm$0.27 &  & & 12.77 \nl
124 & N4552 & $\rm S0_{1}$(0) &.14&13.10$\pm$0.68 & 13.30$\pm$0.30 & & 13.42 \nl
126 & N4567/N4568 & Sc(s)II-III+Sc(s)III&.01  & 11.57$\pm$0.26 & 12.15$\pm$0.10 & & 12.22 \nl
127 & IC3576 & SBdIV &-.03& 12.63$\pm$0.25 & & &  \nl
129 & N4569 & Sab(s)I-II &.08& 11.38$\pm$0.21 & 12.46$\pm$0.13 & & 11.39 \nl
131 & N4571 & Sc(s)II-III & .07&11.86$\pm$0.25 & & &\nl
132 & IC3591 & IB & -.03&12.99$\pm$0.31 &  & &\nl
136 & VCC 1725 & SmIII/BCD &-.02& 13.10$\pm$0.27 & & & \nl
137 & U7795 & Sdm  & -.05&13.62$\pm$0.51 & &  &  \nl
138 & N4579 & Sab(s)II &.14& 12.03$\pm$0.32 & 12.84$\pm$0.18 & & 12.37 \nl
144 & IC3617 & SmIII/BCD&-.01 & 13.17$\pm$0.32 & & & 14.54 \nl
146 & N4595 & Sc(s)II.8 &.02& 12.57$\pm$0.42 & & & 14.39\nl
147 & N4596 & SBa & -.03&12.50$\pm$1.04 & & & 14.64 \nl
155 & KARA 68181 & & .02&13.04$\pm$0.32 & & &  \nl
168 & N4634 & Sc (on edge)& .03 & 12.54$\pm$0.44 & & 13.5 & \nl
170 & N4639 & SBb(r)II &  .05 &11.71$\pm$0.24  & & 12.4 & 12.72 \nl
171 & N4649/N4647 & $\rm S0_{1}$(2)+Sc(rs)III&.04&11.40$\pm$0.20 & & 12.0 & 11.87 \nl
172 & N4651 & Sc & .03&10.97$\pm$0.18 & & & \nl
174 & N4654 & SBc(rs)II& .05&10.86$\pm$0.18 & & 11.2 & 11.67\nl
178 & IC3742 & SBc(s)II &.06& 12.72$\pm$0.55 & & & \nl
187 & N4689 & Sc(s)II.3 &.04& 11.59$\pm$0.31 & & 12.2 & \nl
\enddata
\end{deluxetable}

\newpage

\newpage

\begin{deluxetable}{rrcrrrrrr}
\tablecaption{Galaxies towards the Virgo cluster-Multispectral data}
\tablehead{\colhead{N} & \colhead{Name} & 
\colhead{T-Type} & \colhead{FAUST}& \colhead{FI(HI)} & 
\colhead{Ir [60]} & \colhead{Ir [100]} & 
\colhead{v$_{r}$} & \colhead{Bc$_{tot}$}}
\startdata
2  & N4178 & 5 & 11.61$\pm$0.19 & 56.2 & 2.11 & 8.08 & 381& 11.89\nl
12 & IC3111 & 3 &  13.03$\pm$0.32 & .24540 & .67 & 0.72   &14755  & 14.78\nl
15 & N4246 & 5 & 12.83$\pm$0.35 & 21.5 &  &  & 3724 & 13.38  \nl
17 & N4254 & 5 & 9.80$\pm$0.17 & 105.6 & 25.3 &  76.01  &2470& 10.30 \nl
20 & IC0776 & 6 & 13.42$\pm$0.31 & 10.2 & &  & 2464 & 14.30 \nl
23 & N4276 & 5  & 12.90$\pm$0.31 &  4.66 & .48 & 1.64  & 2628 & 13.25 \nl
26 & N4294 & 5 & 11.53$\pm$0.17 &  & .28 & 9.40  &  359  & 12.51  \nl
28 & N4299 & 6 & 11.54$\pm$0.17 & 13.5 & 2.64 & 8.09 &  221&  12.86 \nl
29 & N4298 & 5 & 12.01$\pm$0.31 & 12.9 & 5.09 & 21.27 & 1133 &  11.94 \nl
31 & N4321 & 5 & 9.99$\pm$0.17 & 50.0 & 18.95 & 56.19 & 1560&  10.01 \nl
32 & N4325 & --5 & 13.73$\pm$0.97 & & .75& 1.77 & 7786  & 14.30\nl
34 & N4330 & 7 & 13.22$\pm$0.35 &  &.63 & 2.78 & 1552  & 13.06 \nl
35 & IC3258 & 5 & 12.32$\pm$0.24 & 3.69 & .49 & .97 & --437 & 13.57\nl
38 & N4351 & 5 & 12.75$\pm$0.38 &  4.17 & .71 & 2.01 & 2320 & 13.02\nl
39 & IC3268 & 5,9 & 12.60$\pm$0.25 &  8.54 & .69 & 1.65  & 697 & 14.22\nl
41 & N4365 & --5 & 13.98$\pm$0.61 & $\leq$2.65 &  &  & 1240 & 10.52  \nl
42 & VCC 740 & 9 & 14.10$\pm$0.60 &           &  &  & 874 & 15.70\nl
43 & N4371 & --1 & 13.80$\pm$0.44 &  $\leq$3.02  &  &  & 941& 11.77  \nl
44 & N4374 ? & --5 & 13.30$\pm$0.63 &  $\leq$2.19 & .50 & .98 & 1033 &  9.96\nl
45 & N4380 & 2 & 12.81$\pm$0.38 &  3.02 & .63 & 3.12 & 963  & 12.66  \nl
46 & N4383 &  & 11.71$\pm$0.21 &  44.2 & 7.95 & 12.32 & 1694 &  12.65 \nl
47 & IC3322A & 5 & 12.93$\pm$0.24 & 22.7 & 1.71 & 5.39 & 1001 & 13.55  \nl
48 & N4390 & 4 & 12.73$\pm$0.40 & 7.07 & .75 & 1.88  & 1118 & 13.30\nl
49 & N4388 & 2 & 12.13$\pm$0.27 &  6.46 & 10.24 & 18.1  & 2535 &  11.65 \nl
50 & N4396 & 5 & 12.34$\pm$0.45 & & 1.15 & 3.89 & --155 & 13.02 \nl
51 & N4406 & --1, --5 & 12.94$\pm$0.79 & $\leq$2.51  & &  & --221 & 9.72  \nl
52 & N4413 ? & 4 & 11.85$\pm$0.31 &  3.39  & 1.04 & 3.19  & 96 & 12.14  \nl
53 & N4411A & 5 & 11.32$\pm$0.23 & 7.50 & & & 1280 & 13.41  \nl
54 & N4411B & 5 & 11.85$\pm$0.17 & 21.88 & .40 & 1.78  & 1269  & 12.91\nl
55 & N4416 & 5 & 12.41$\pm$0.20 &  4.79 & .93 & 2.70 & 1380  & 13.14 \nl
57 & VCC 975 ? & 6  & 12.27$\pm$0.17 & 17.7  & &  &  932 & 13.58 \nl
58 & N4424 & 1 & 14.16$\pm$0.59 & 2.82 & 3.31 & 5.92 & 432 &  12.34  \nl
59 & IC3365 & 6 & 12.71$\pm$0.64 & 3.52  &  &  &  2336 & 14.33  \nl
63 & N4432 & 5 & 11.78$\pm$0.21 & 5.1 & .25 & 1.27 & 6403  &14.70  \nl
66 & N4438 & 3 & 12.18$\pm$0.39 & 7.76 & 3.76 & 11.27  &  86 &  10.92 \nl
69 & VCC 1091 ? & 4 & 12.69$\pm$0.23 & 16.13 & .38 & .82 & 1117 & 14.6  \nl
71 & N4451 & 5 & 13.77$\pm$0.46 & 2.68 & 1.68 & 4.56   &  860 & 13.29 \nl
78 & IC3414 & 5 & 13.11$\pm$0.31 & 3.63 & .23 & .73  &  525 & 13.94 \nl
81 & N4470 & 5,13 & 12.13$\pm$0.19 &  8.32 & 1.82 & 4.46 & 2360  & 13.02 \nl
83 & N4472 & --5, --1 & 12.17$\pm$0.19 &  $\leq$0.72 & &  & 997  & 9.37 \nl
88 & N4486 & --5 & 11.84$\pm$0.30 &  & .39 & 1.02  & 1292 & 9.50 \nl
90 & IC3446 & 9 & 13.39$\pm$1.28 & & &   & 1251 & 14.87 \nl
91 & N4489 & --1 & 12.03$\pm$0.26 &  & 1.24 & 4.03 & 1507 & 12.81 \nl
94 & IC0797 & 5 & 12.95$\pm$0.65 & 3.94 & .74 & 2.18 & 2100 & 13.46 \nl
95 & N4501 & 4 & 10.92$\pm$0.18 & 33.88 &  &  & 2321  & 10.27 \nl
100 & IC3476 & 5 & 11.89$\pm$0.38 & 4.57 & 1.35 & 3.24 & --169 & 13.10  \nl
106 & Q 1230+0947 &  & 13.92$\pm$1.30 &  & & & &\nl
107 & N4519/4519a & 5,0 & 11.13$\pm$0.23 & 51.29 & 3.88 & 6.65  & 1229  & 12.34\nl
109 & N4522 & 5,3 & 12.26$\pm$0.45 &  7.08 & 1.30 & 4.20 & 2316  & 12.99  \nl
110 & N4523 & 7 & 11.69$\pm$0.30 & 25.9 & .38 & 1.30 & 262 & 14.33 \nl
111 & VCC 1525 & 4 & 14.17$\pm$0.63 &  & .36 & .83   & 11382 & 14.81\nl
113 & N4532 & 9 & 10.95$\pm$0.16 & 42.3 & 8.95 & 16.21  & 2010& 12.30  \nl
114 & N4535/4535a & 3?,5 & 10.18$\pm$0.16 & 89.13 & 6.88 & 23.61 & 1962& 10.50   \nl
118 & IC3528 & 4 & 13.08$\pm$0.36 & 1.4  & 1.53 & 5.52 & 13830 & 15.20 \nl
121 & VCC 1600 ? & --5  & 13.32$\pm$0.52 &  & & & & 19.00  \nl
123 & N4548 & 3 & 11.76$\pm$0.27 & 9.25& 1.47 & 9.44  & 550  & 10.89\nl
124 & N4552 & --1 & 12.82$\pm$0.68 & $\leq$2.51 & & & 322 & 10.59 \nl
126 & N4567 or N4568 & 5 & 11.55$\pm$0.26 & 21.38  & 16.91 & 47.88 & 2186 & 12.05+ \nl
    &     &   &                 &       &       &     &  & 11.67 \nl
127 & IC3576 & 7 & 12.63$\pm$0.25 & 14.5& & & 1076 & 13.70\nl
129 & N4569 & 2 & 11.22$\pm$0.21 &  12.3  & 7.56 & 23.66 & -223 & 10.18 \nl
131 & N4571 & 5 & 11.72$\pm$0.25 & 12.8 & 1.10 & 5.78 & 343  & 11.75  \nl
132 & IC3591 & 10 & 12.99$\pm$0.31 &  & & & 1635  & 14.40 \nl
136 & VCC 1725 & 9 & 13.10$\pm$0.27 & & & & 1065  &14.51 \nl
137 & U7795 & 8 & 13.62$\pm$0.51  & 4.82 & &  & 61 & 14.54  \nl
138 & N4579 & 2 & 11.75$\pm$0.32 &  8.91 & &  & 1520  & 10.34  \nl
144 & IC3617 & 9 & 13.17$\pm$0.32 & 6.3  & .27 & .64  & 2079  & 14.37\nl
146 & N4595 & 5 & 12.53$\pm$0.42 &  & .91 & 2.81  & 630 & 12.89  \nl
147 & N4596 & 1 & 12.50$\pm$1.04 & $\leq$3.71 & .49 & 1.28 & 1870  & 11.35 \nl
155 & KARA 68181 &  & 13.00$\pm$0.32 &  & &  &  &\nl
168 & N4634 & 5 & 12.48$\pm$0.44 &  & 4.04 & 11.42 & 144  & 13.13\nl
170 & N4639 & 3 & 11.61$\pm$0.24 & 15.8 & 1.41 & 4.63 & 990 & 12.19    \nl
171 & N4649 or N4647 & --1, 5 & 11.32$\pm$0.20 & $\leq$6.88 or 2.1& 5.33 & 15.42  & 1095  & 9.77+\nl
  &     &   &                 &       &       &     &  & 11.90 \nl
172 & N4651 & 3? & 10.17$\pm$0.18 & 59.9  & 5.45 & 15.57  & 800 & 11.36  \nl
174 & N4654 & 5 & 10.76$\pm$0.18 &  68.3 & 13.08 & 34.91 & 1044 & 11.05  \nl
179 & IC3742 & 5 & 12.60$\pm$0.55 & 4.24  & .25 & .94  & 872 & 14.00 \nl
187 & N4689 & 5 & 11.51$\pm$0.31 &  6.0 & 2.55 & 10.99  & 1616 & 11.56 \nl
\enddata

\tablecomments{FI(HI) is the HI flux integral, 
in units of Jy km s$^{-1}$.}

\end{deluxetable}
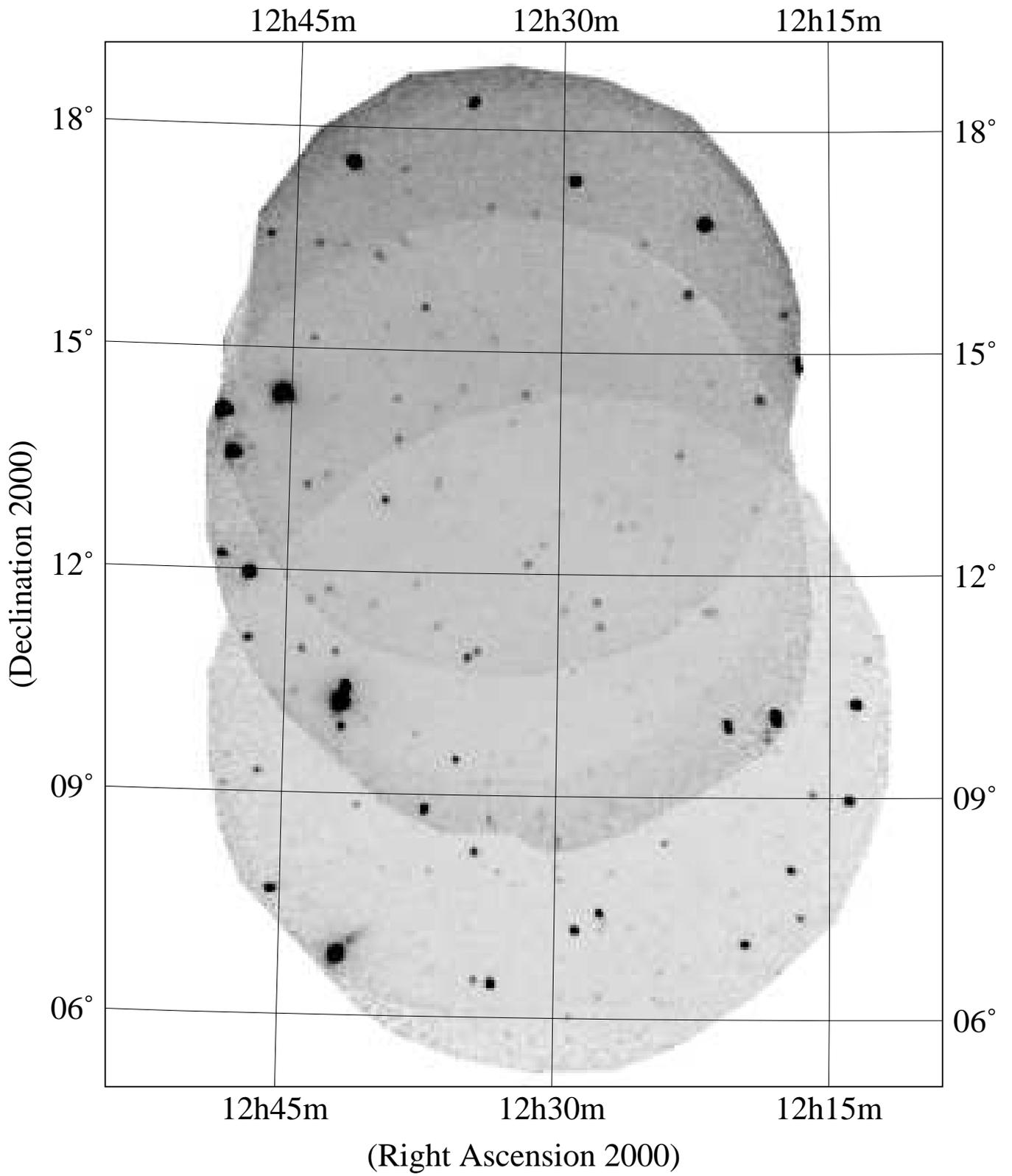
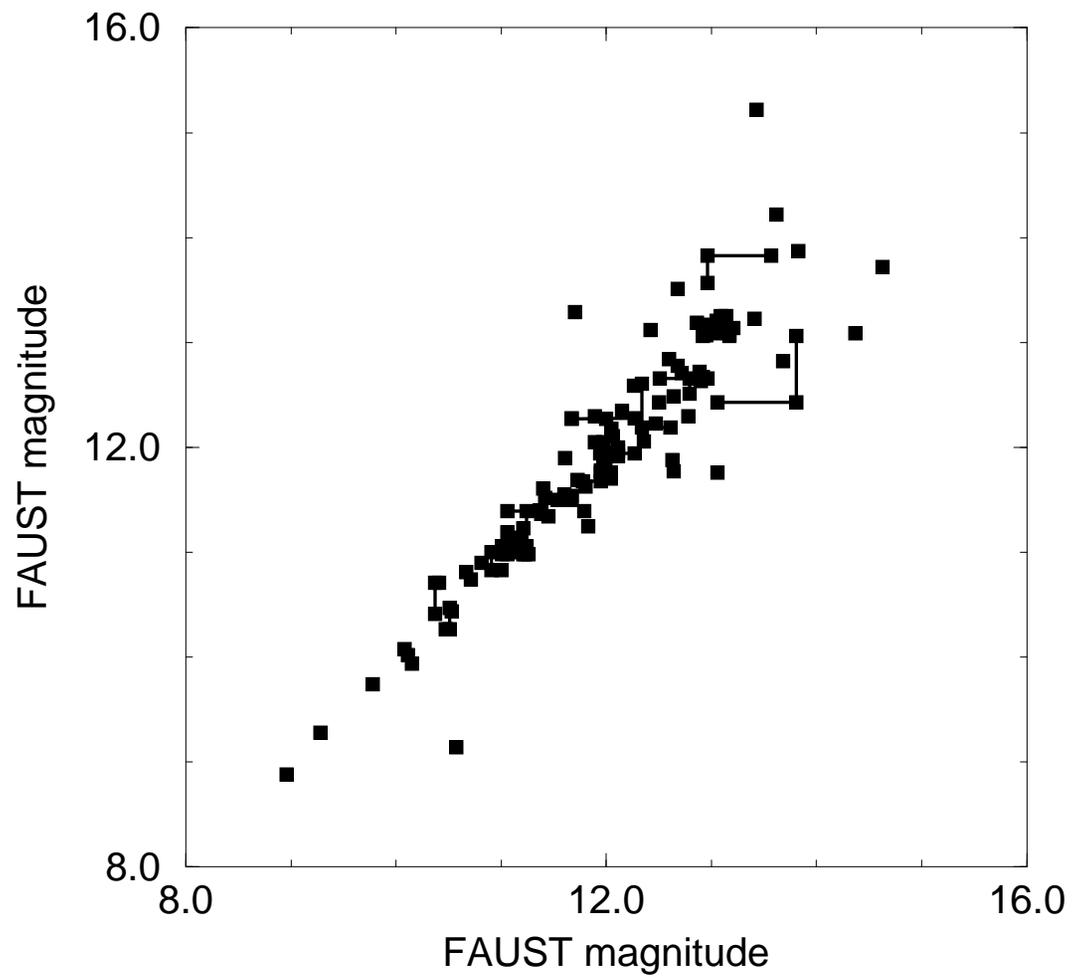
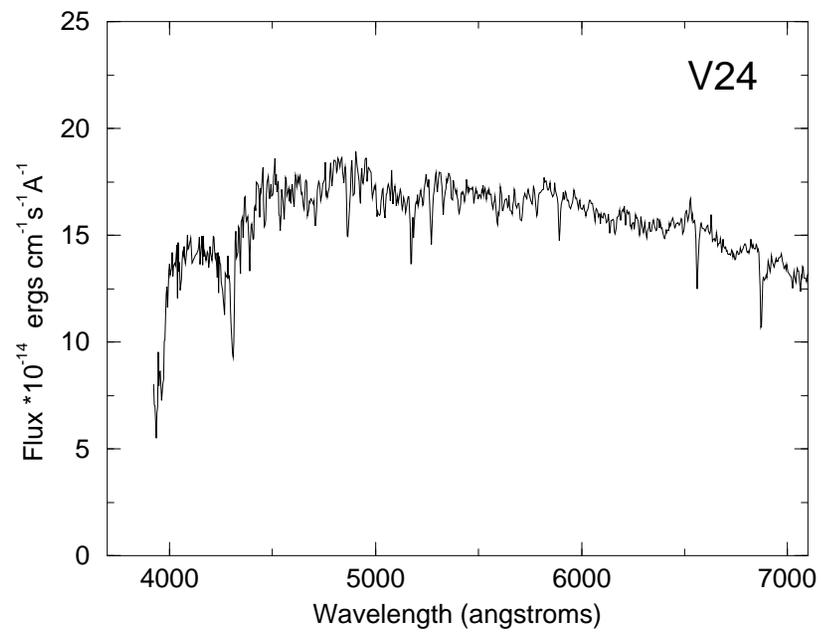
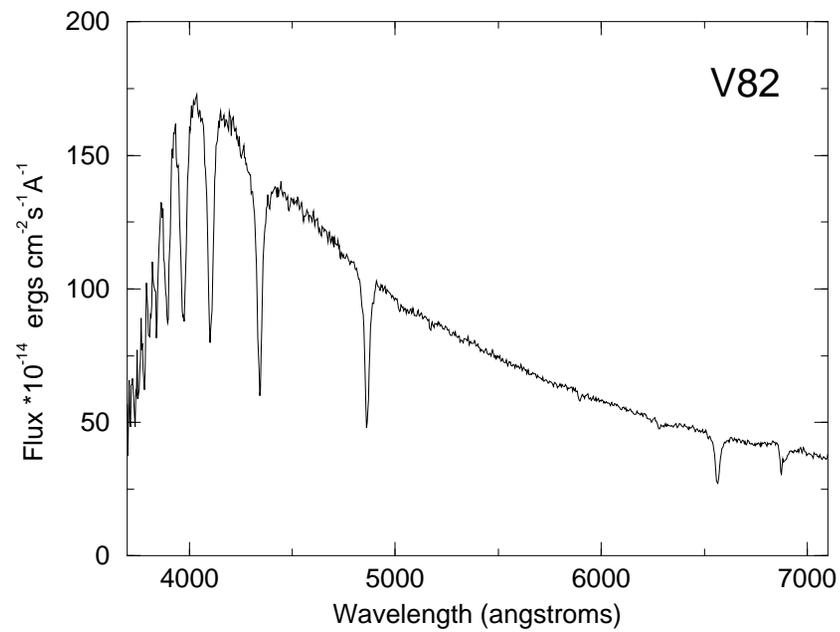
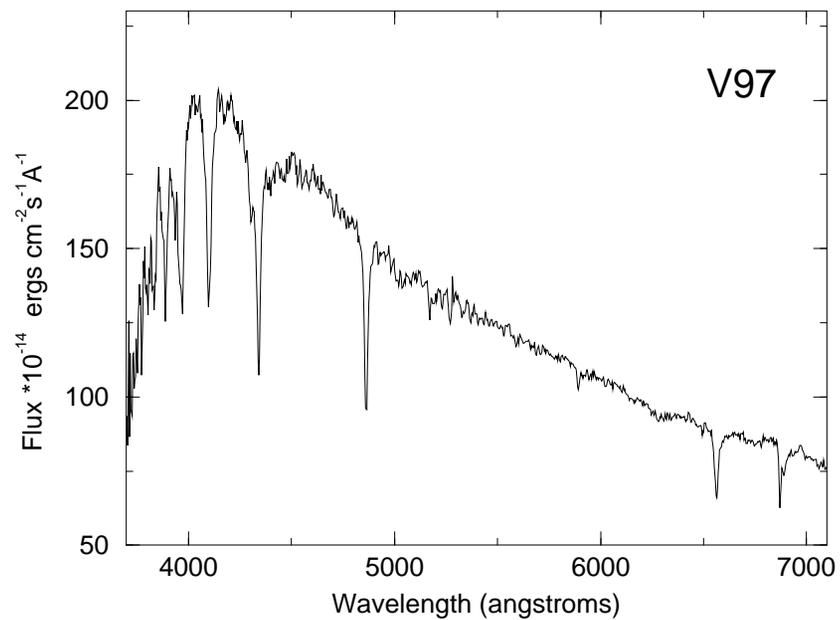
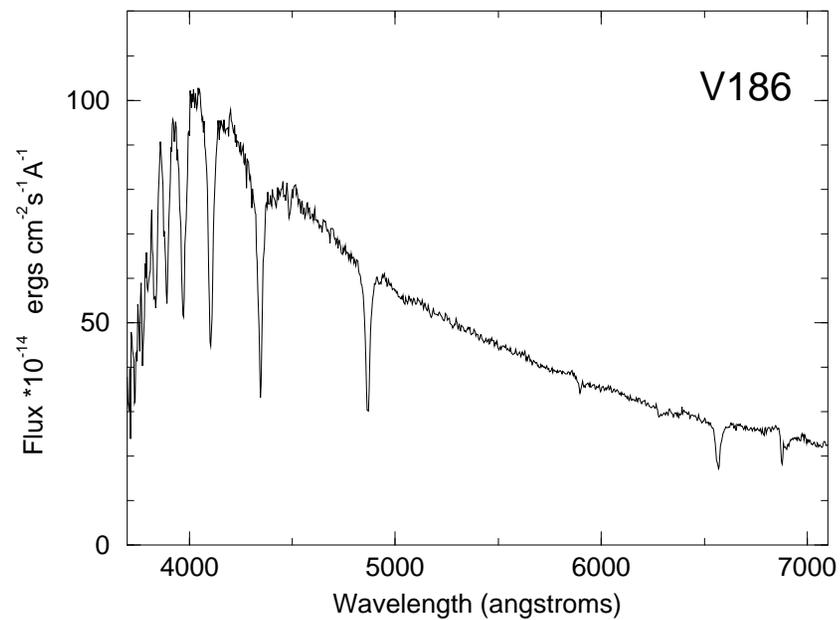
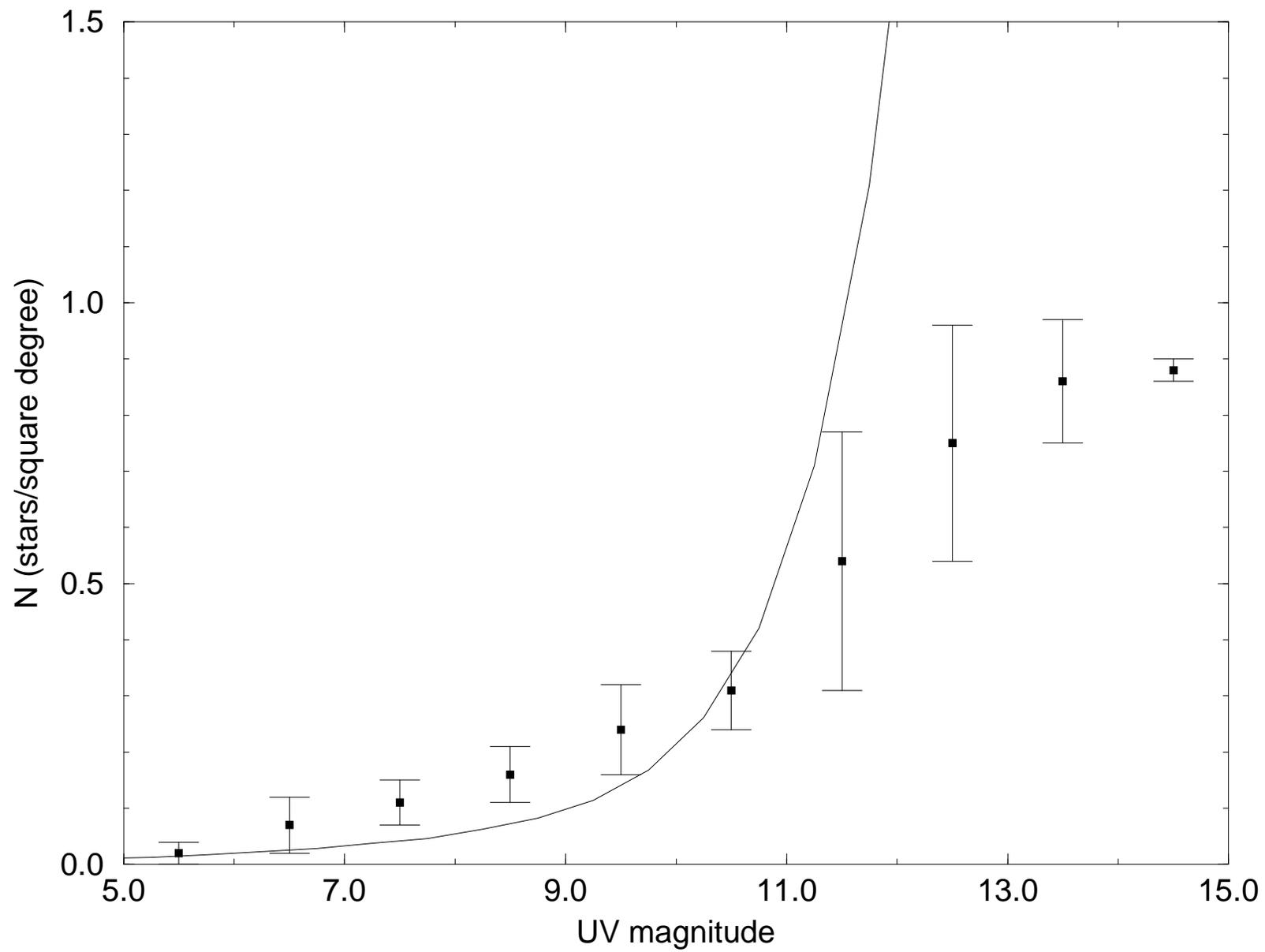
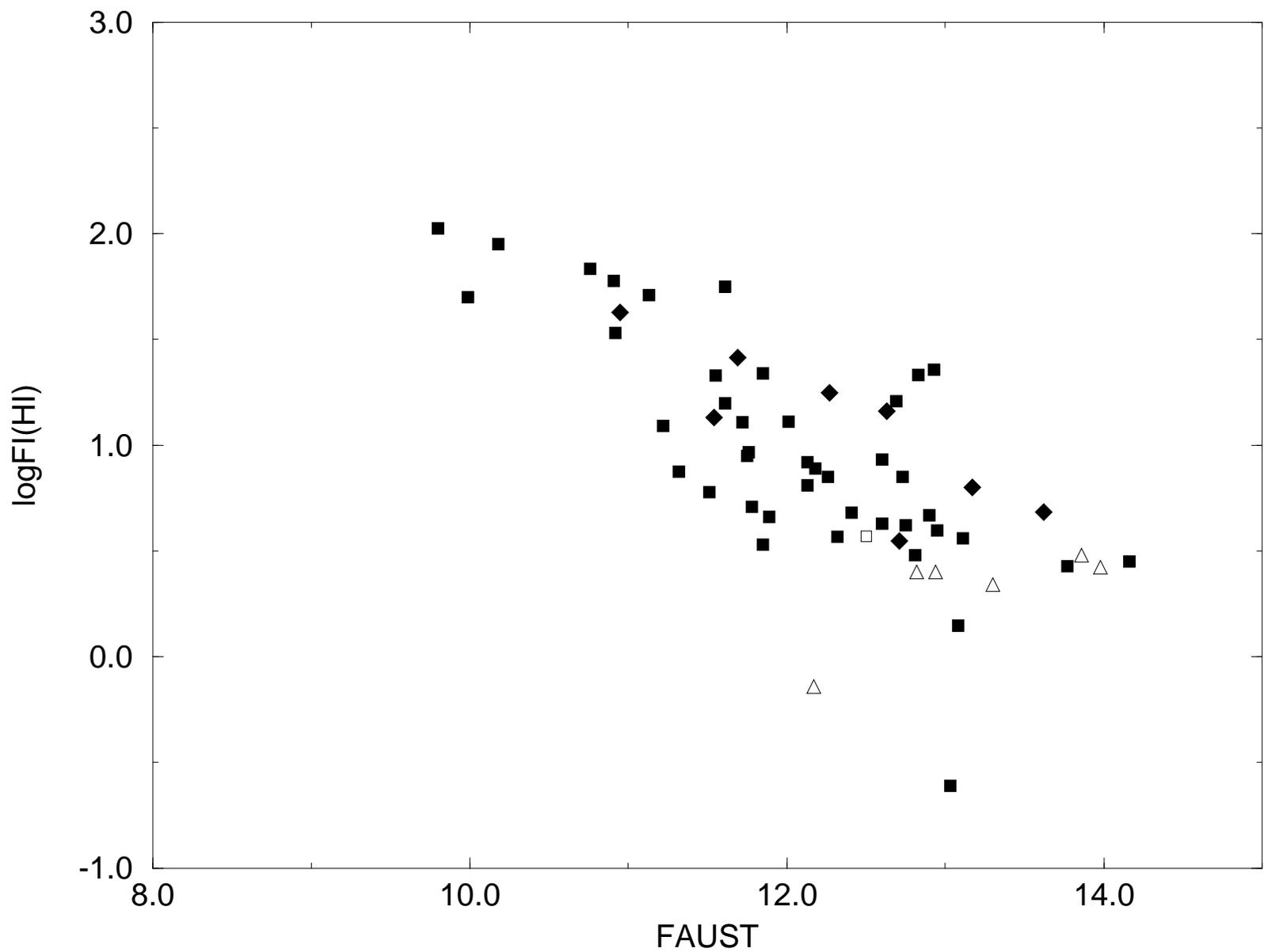
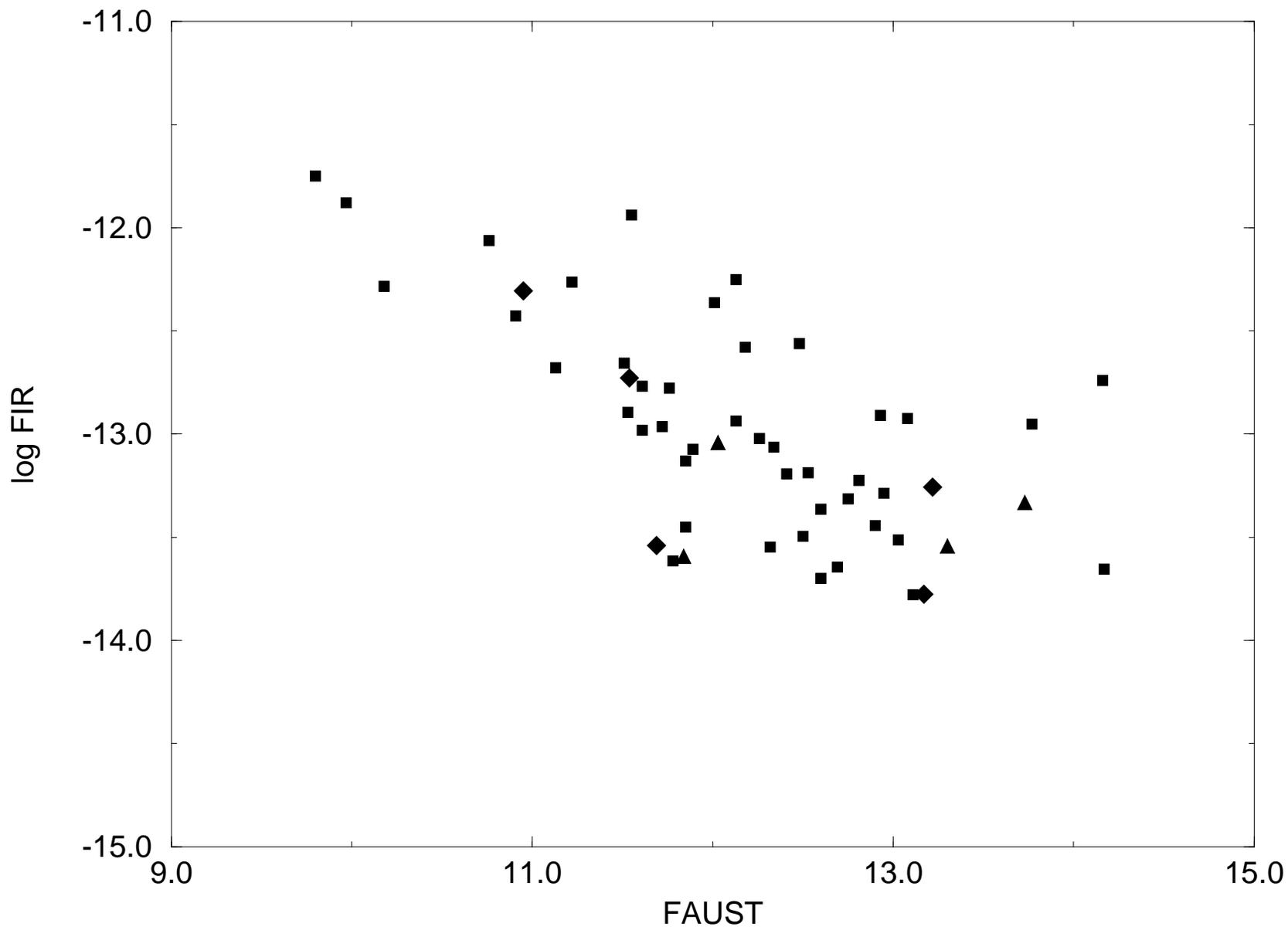
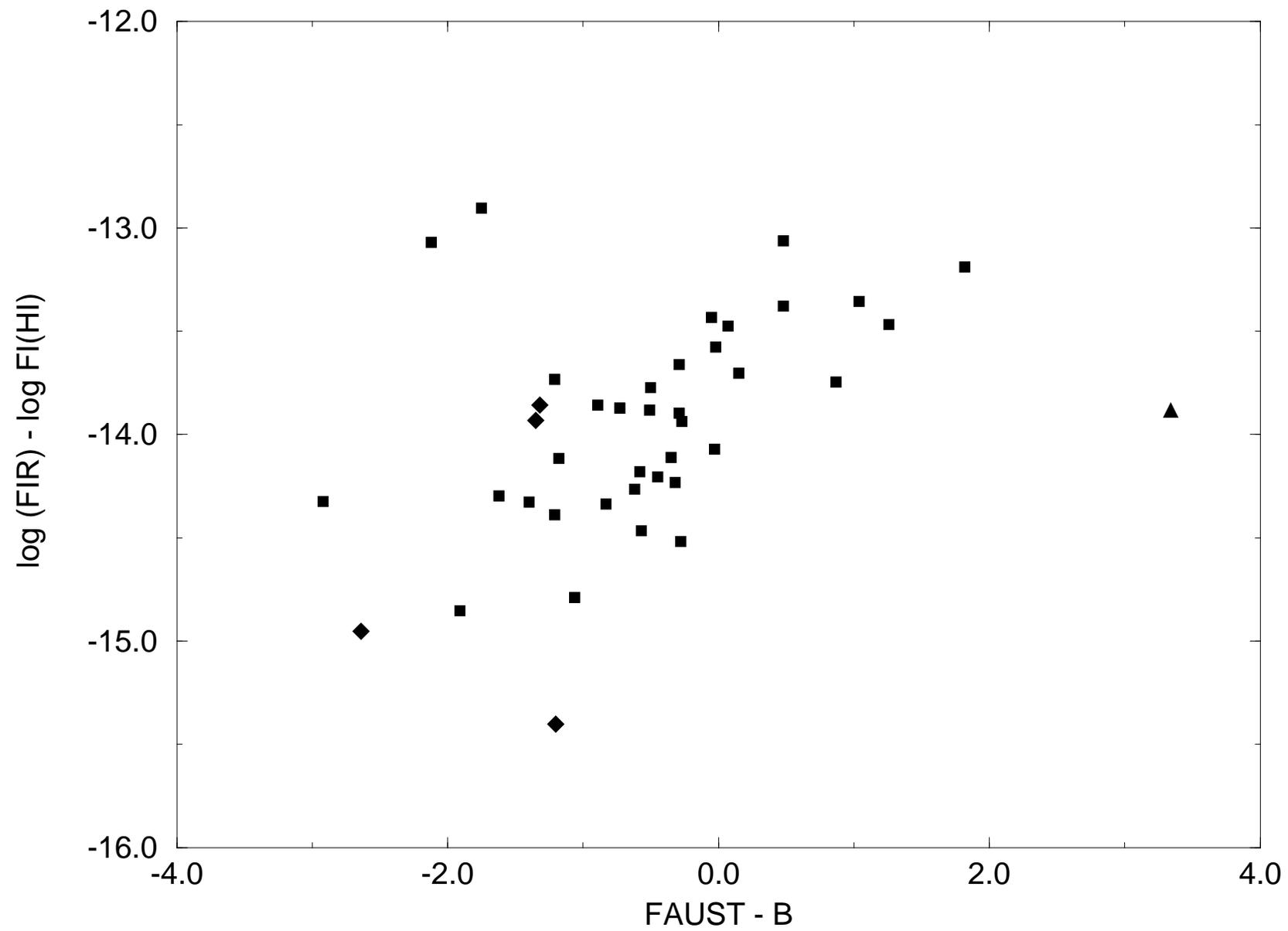
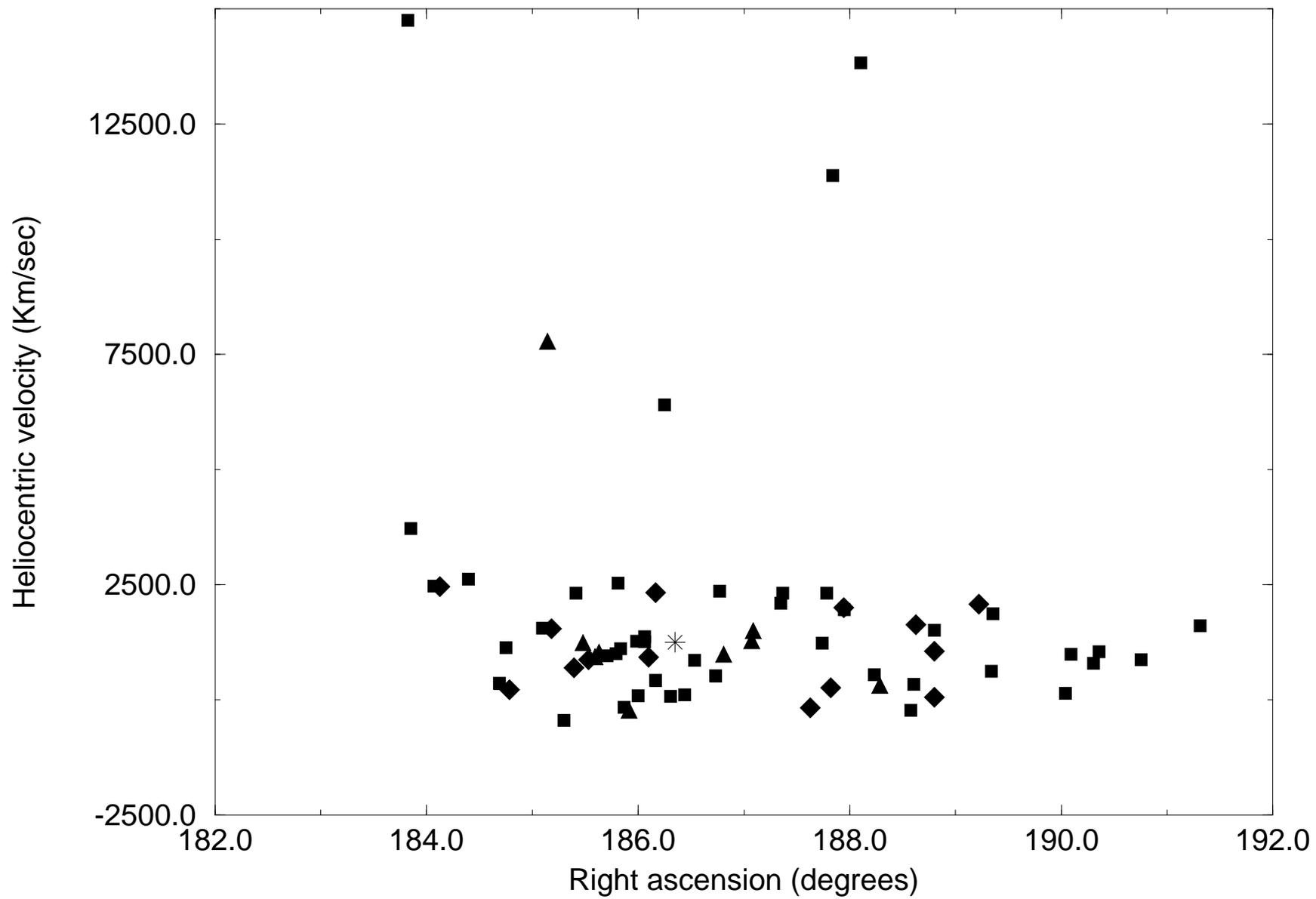
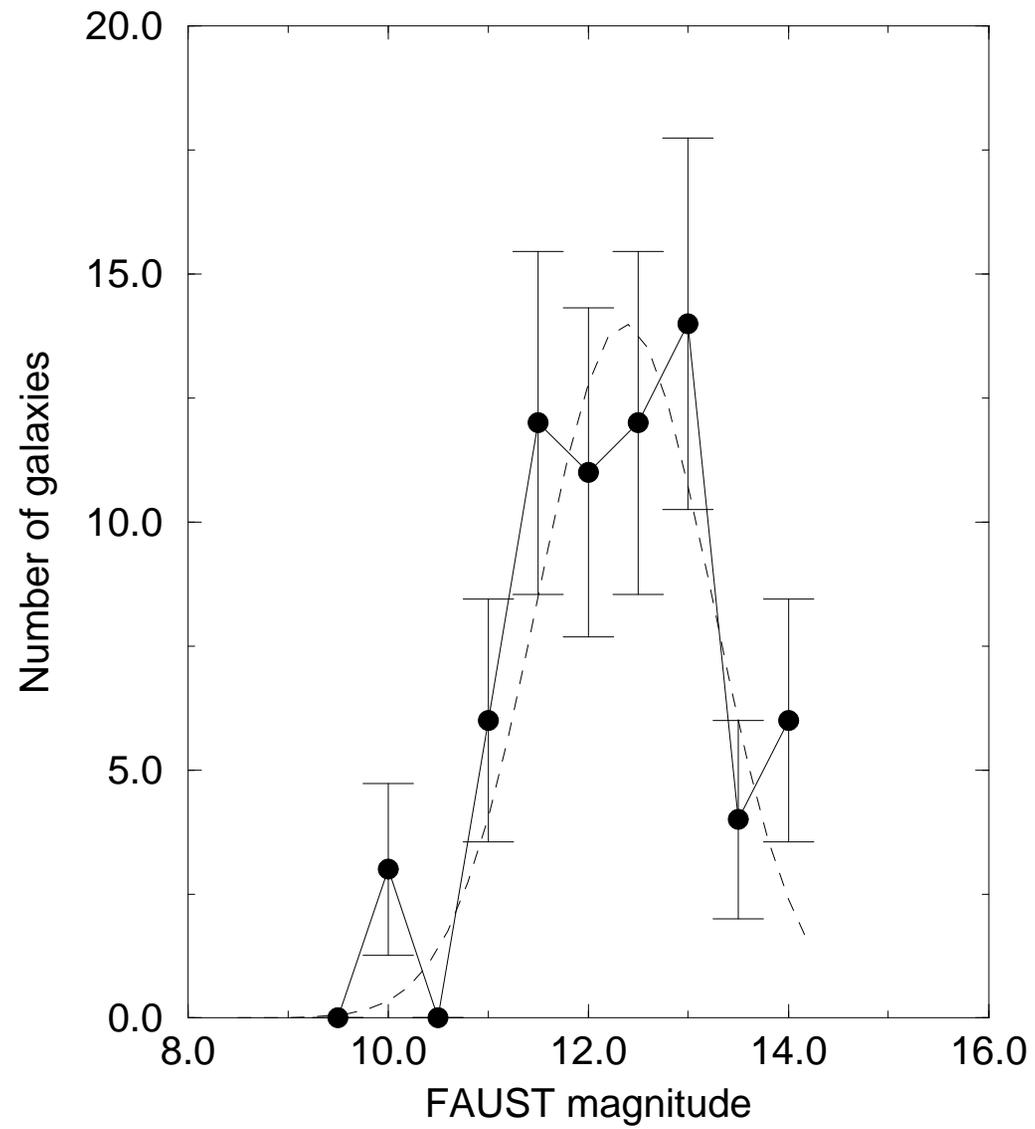
\end{document}